\documentclass[a4paper,fleqn,usenatbib]{mnras}

\usepackage[T1]{fontenc}
\usepackage{ae,aecompl}

\usepackage{graphicx}	
\usepackage{amsmath}	
\usepackage{amssymb}	

\title[The VMC Survey. XXV. The structure of the SMC]{The VMC Survey. XXV. The 3D structure of the Small
  Magellanic Cloud from Classical Cepheids}

\author[Vincenzo Ripepi et al.]{
Vincenzo Ripepi,$^{1}$\thanks{E-mail: ripepi@oacn.inaf.it (VR)}
Maria-Rosa L. Cioni,$^{2,3}$
Maria Ida Moretti,$^{1}$
Marcella Marconi,$^{1}$
\and  
Kenji Bekki,$^{4}$ 
Gisella Clementini,$^{5}$ 
Richard de Grijs,$^{6,7}$
Jim Emerson,$^{8}$
\and  
Martin A. T. Groenewegen $^{9}$
Valentin D. Ivanov,$^{10,11}$
Roberto Molinaro,$^{1}$ 
\and   
Tatiana Muraveva,$^{5}$ 
Joana M. Oliveira$^{14}$
Andr\'es E. Piatti,$^{12,13}$
Smitha Subramanian,$^{6}$ 
\and   
Jacco Th. van Loon$^{14}$ 
\\
$^{1}$INAF-Osservatorio Astronomico di Capodimonte, via Moiariello 16, 80131, Naples, Italy\\
$^{2}$ Leibniz-Instit\"{u}t f\"{u} Astrophysik Potsdam, An der Sternwarte 16, 14482 Potsdam, Germany \\
$^{3}$ University of Hertfordshire, Physics Astronomy and Mathematics, College Lane, Hatfield AL10 9AB, UK \\
$^{4}$ ICRAR M468 The University of Western Australia 35 Stirling Hwy, Crawley Western Australia 6009, Australia\\
$^{5}$ INAF-Osservatorio Astronomico di Bologna, Via Gobetti 93/3, 40129, Bologna, Italy\\
$^{6}$  Kavli Institute for Astronomy \& Astrophysics and Department of Astronomy, Peking University, Yi He Yuan Lu 5, \\
~~~Hai Dian District, Beijing 100871, China \\
$^{7}$ International Space Science Institute--Beijing, 1 Nanertiao, Zhongguancun, Hai Dian District, Beijing 100190, China \\
$^{8}$ Astronomy Unit, School of Physics and Astronomy, Queen Mary University of London, Mile End Road, London E1 4NS, UK \\
$^{9}$ IKoninklijke Sterrenwacht van Belgi\"{e}, Ringlaan 3, B-1180 Brussels, Belgium\\
$^{10}$ European Southern Observatory, Ave. Alonso de Cordova 3107, Vitacura, Santiago, Chile\\
$^{11}$ European Southern Observatory, Karl-Schwarzschild-Str. 2, D-85748 Garching bei M\"{u}nchen, Germany\\
$^{12}$ Observatorio Astron\'omico, Universidad Nacional de C\'ordoba, Laprida 854, 5000, C\'ordoba, Argentina\\
$^{13}$ Consejo Nacional de Investigaciones Cient\'ificas y T\'ecnicas, Av. Rivadavia 1917, C1033AAJ, Buenos Aires, Argentina\\
$^{14}$ Lennard-Jones Laboratories, Keele University, ST5 5BG, UK
}

\date{Accepted XXX. Received YYY; in original form ZZZ}

\pubyear{2017}

\begin{document}
\label{firstpage}
\pagerange{\pageref{firstpage}--\pageref{lastpage}}
\maketitle

\begin{abstract}
The {\it VISTA near-infrared $YJK_\mathrm{s}$ survey of the
    Magellanic System} (VMC) is collecting deep $K_\mathrm{s}$-band
  time--series photometry of pulsating stars hosted by the two
  Magellanic Clouds and their connecting Bridge. Here we
present  $Y,\,J,\,K_\mathrm{s}$ light curves for a sample of 717 Small Magellanic Cloud (SMC) Classical
  Cepheids (CCs). These data, complemented with our previous results
  and $V$ magnitude from literature, allowed us to construct a variety of
  period--luminosity and period--Wesenheit relationships, valid for Fundamental,   
 First and Second Overtone pulsators. These 
 relations provide accurate individual distances to CCs in the SMC
 over an area of more than 40 deg$^2$. Adopting literature relations,
 we estimated ages and metallicities for the majority of the
 investigated pulsators, finding that: i) the age distribution is bimodal,
 with two peaks at 120$\pm$10 and 220$\pm$10 Myr; ii) the
more metal-rich CCs appear to be located closer to the centre of the galaxy.  
Our results show that the three--dimensional distribution of the CCs
in the SMC, is not planar but heavily elongated for more than 25-30 kpc approximately in the
east/north-east towards south-west direction. The young and old CCs in
the SMC show a different geometric distribution. Our data support the current theoretical
scenario predicting a close encounter or a direct collision between
the Clouds some 200 Myr ago and confirm the
presence of a Counter-Bridge predicted by some models. The high
precision three--dimensional distribution of young stars presented in this paper
provides a new testbed for future models exploring the formation
and evolution of the Magellanic System.

\end{abstract}




\section{Introduction}

The Large Magellanic Cloud (LMC) and the Small Magellanic Cloud (SMC)
are gas-rich dwarf irregular galaxy satellites of the Milky Way (MW).  The
Magellanic Clouds (MCs) and the MW represent the closest group of
interacting galaxies, providing the best opportunity to investigate
satellite-satellite and satellite-host galaxy gravitational
interactions \citep[see e.g.][]{Donghia2016}.
Moreover, the MCs are fundamental benchmarks in the context of stellar
populations and galactic evolution studies \citep[see, e.g.,][and
references therein]{Harris2004,Harris2009,Ripepi2014a}.  Indeed, they
are fairly close \citep[$D \sim50-60$
kpc;][]{deGrijs2014,deGrijs2015}, host stellar populations of diverse 
ages/metallicities, and their morphologies were notably affected by
the dynamical interaction between them and with the MW.  

There are clear signatures that the SMC is interacting with both the LMC and the MW.  In particular, the MCs are
connected by a Magellanic Bridge (MB) traced by \ion{H}{i} gas but also including a
significant stellar content \citep[e.g.][]{Irwin1985,Harris2007}.  Like
the Magellanic Stream (MStr), an H{\sc i} structure embracing the MCs and
extending over a large region of the sky, the MB may be the
signature of the MCs' mutual gravitational interaction and/or the
impingement of the MW \citep[e.g.][]{Putman1998,Hammer2015}.
Furthermore, the SMC Wing or Shapley Wing \citep{Shapley1940},
extending asymmetrically towards the LMC, could be
the result of tidal interaction(s). Additionally, the asymmetric and
elongated SMC shape, especially when traced by the young population
reveals the strength of the gravitational forces acting between the two
Clouds. 

Very recent investigations of MCs' outskirts revealed a perhaps 
more complex structure of the whole Magellanic System (MS), including either
relics of its formation or the results of strong tidal
interaction between the MCs. Indeed: i)   
a number of ultra--faint dwarf galaxies, recently discovered in the Dark
Energy Survey (DES), might be associated with the
LMC \citep[][]{Deason2015,Drlica2015,
  Drlica2016,Koposov2015,Jethwa2016,Martin2016,Sales2017}; 
ii) a stellar protuberance extending by about 10 kpc was discovered by
\citet{Mackey2016} within the LMC's tidal radius; iii)
\citet{Belokurov2017} used {\it Gaia} satellite data release 1
\citep[][]{Gaia2016a,Gaia2016b} to
reveal that the LMC and SMC tidal arms are stretched towards each other 
to form an almost continuous (new) stellar bridge; iv) a stellar overdensity  
 called Small Magellanic Cloud Northern Over-Density (SMCNOD) was
 discovered 8$\degr$ north (N) of the SMC centre by \citet{Pieres2017} 
on the basis of DES, Survey of the MAgellanic Stellar
History (SMASH) and MAGellanic SatelLITEs Survey (MagLiteS) data. 
These new findings show that perhaps the inventory of the MS is still
incomplete.

In this context, recent studies based on proper motion estimates, using Hubble Space
Telescope (HST) and {\it Gaia} satellite data, suggest that the MCs are
at their first passage of the MW \citep[][and references
therein]{Besla2007,Kallivayalil2013,vandermarel2016}. This 
suggests that both the MStr and MB are the result of strong interactions
between the two Clouds before they interacted with the
MW. Furthermore, 
simulations by \citet{Diaz2012} predict close encounters
$\sim$2 Gyr and $\sim$200 Myr ago, leading to the formation of the MStr
and MB, respectively. The same models also predict the existence of a
Counter-Bridge (CB) in the opposite direction to the MB.  Other models
\citep{Besla2012} reproduce the formation of the MStr, the off-center
bar of the LMC and the formation of the MB, this last under the hypothesis of an
impact between the MCs at an epoch between 100 and 300 Myr ago.

Focussing on the SMC, it is commonly thought that this galaxy is composed of two components:
i) young stars and \ion{H}{i} gas forming a disk; ii) old
and intermediate-age populations more smoothly distributed in a
spheroid or ellipsoid \citep[see e.g.][among 
others]{Caldwell1986,Gardiner1996,Cioni2000,Zaritsky2000,Maragoudaki2001,Stanimirovic2004,
  Harris2006,Bekki2008,Evans2008,Glatt2008,Gonidakis2009,Haschke2012,
  Subramanian2012,Subramanian2015,Rubele2015}.  A common finding among
many of the quoted works is that the north-eastern region of the
bar/disk is closer to us than its south-western part \citep[as early-on
recognized by][]{Welch1987,hatzi89}.  Moreover, the SMC shows a
considerable line-of-sight (LOS) depth whose precise quantitative
extent depends on the methods and tracers used for the measure spanning
a few kpc to more than 20 kpc \citep[see, e.g.][for a
review]{deGrijs2014}.

Our view of the SMC and its interaction with its neighbour has changed
significantly in recent years. \citet{Nidever2013} used
Washington photometry of Red Clump (RC) stars in eight MOSAIC@CTIO-4m
fields to identify a stellar structure significantly closer to us 
(D$\sim$55 kpc) than the main body of the SMC (placed at D$\sim$67 kpc
in that work), in its eastern part (around 4.2 kpc from SMC
centre). They interpret this structure as a component which was
tidally stripped 
during the last interaction with the LMC ($\sim$200 Myr ago). Using
synthetic Colour Magnitude Diagram (CMD) fitting technique,
\citet{Noel2015} suggest that the intermediate-age population in the
region of the MB closer to the SMC was tidally stripped from the
galaxy, as its properties are similar to those of the intermediate-age
stellar populations in the inner regions of the galaxy.  From a
kinematic investigation of Red Giants in the SMC, \citet{Dobbie2014}
found the presence of tidally stripped stars associated with the MB.

More recently, on the basis of the {\it VISTA\footnote{Visible and
    Infrared Survey Telescope for Astronomy} near-infrared
  $YJK_\mathrm{s}$ survey of the Magellanic Clouds system} \citep[VMC;
][see below]{Cioni2011}, \citet{Subramanian2017} identified a foreground
population ($\sim$11.8$\pm$2.0 kpc in front of the main body), whose
most likely explanation is tidal stripping from the SMC. Moreover, they
identify the inner region ($\sim$2-2.5 kpc from the centre) from where
the signatures of interactions start becoming evident, thus supporting
the hypothesis that the MB was formed from tidally
stripped material from the SMC.

As for the detailed three--dimensional structure of the SMC, the classical
pulsating stars, RR Lyrae and Classical Cepheids (CCs), have been widely
used in the literature as distance indicators and tracers of the old
(age $>$10 Gyr), and young (age typically $\sim$ 50-500 Myr) populations,
respectively. Concerning RR Lyrae, very recent investigations based on
both near-infrared (NIR) time--series data (VMC survey)
and optical OGLE\,IV survey data \citep{Udalski2015}, found that the
distribution of old pulsators in the SMC has an ellipsoidal shape, with a significant
LOS depth ($\sim$ 1-10 kpc) devoid of any particular
substructure, though with some asymmetry in the eastern-southeastern
region of the SMC (i.e. roughly in the direction of the LMC) which 
appears to be closer to us \citep[][]{Subramanian2012,Deb2015,Jacy2017,Muraveva2017}.

Using {\it Spitzer} mid-infrared time--series photometry of 92 bright
CCs in the SMC \citet{Scowcroft2016} confirmed that the galaxy is
tilted and elongated, with its eastern side  up to 20 kpc closer
than the western one. In addition they suggested that the investigated
CCs are not distributed on a disk, but rather have a cylindrical
shape. The \citet{Scowcroft2016}'s results also seem to support the
hypothesis of a direct collision between the MCs to explain the fact
that the objects closer to each Cloud when projected on the sky are
also physically closer to each other.  More recently, \citet{Jacy2016}
presented a thorough investigation of the three--dimensional structure of
the MCs based on OGLE\,IV optical time--series data for thousands of CCs in
both clouds. They confirmed \citet{Scowcroft2016}'s results, i.e. the
SMC CCs do not have a disk-like distribution, but their structure can be
described as an extended ellipsoid. In addition they identified two
large ellipsoidal off-axis structures. The northern one is located
closer to the Sun and it is younger, whereas the south-western structure
is more distant and older. \citet{Jacy2016}, confirming \citet{Subramanian2015}'s
earlier results, found that the age distribution of the CCs
in the SMC is almost bimodal with two peaks at 110 Myr and 220
Myr \citep[an average error of $\sim$20 Myr is
provided by][]{Subramanian2015}. According to their results, younger pulsators are located in the
closer region of the SMC whereas older ones are more distant. These
authors also associate nine CCs with the MB, finding that all of them
are younger than $\sim$300 Myr, in agreement with current theories
about the formation of the MB, which foresee that young stars in MB
are the result 
of in-situ star formation from the tidally stripped material from the SMC. 

In this paper we exploit the $YJK_\mathrm{s}$ time--series
photometry collected in the context of the VMC survey for a sample of
CCs in the SMC including 97.5\% (see below) of the known  pulsators of
this class in this galaxy. The
techniques adopted (template fitting) for the photometry and the
average magnitudes for 4172 SMC CCs were presented
by \citet[][Paper I]{Ripepi2016}. The present work enlarges the sample
of investigated CCs by 15\% (see next Section). In Paper I we adopted our
high-precision photometry (complemented with literature $V$ magnitudes) 
to construct a variety of period--luminosity ($PL$), period--luminosity--colour ($PLC$) and
  period--Wesenheit ($PW$) relationships, valid for Fundamental (F), 
 First Overtone (1O) and Second Overtone (2O) pulsators (the first relation to
 date for this mode of pulsation). The $PW$ was used to
 estimate the distance to the SMC relative to the LMC and, in turn, the absolute
  distance to the SMC. We found $\Delta\mu=0.55\pm0.04$ mag for the
  relative distance and $\mu_{\rm SMC}=19.01\pm0.05$ mag or $\mu_{\rm
    SMC}=19.04\pm0.06$ mag for the absolute value. The two estimates rely on two
  different distance measurements to the LMC: accurate CC and eclipsing Cepheid
  binary data, respectively. 

\begin{figure}
	\includegraphics[width=\columnwidth]{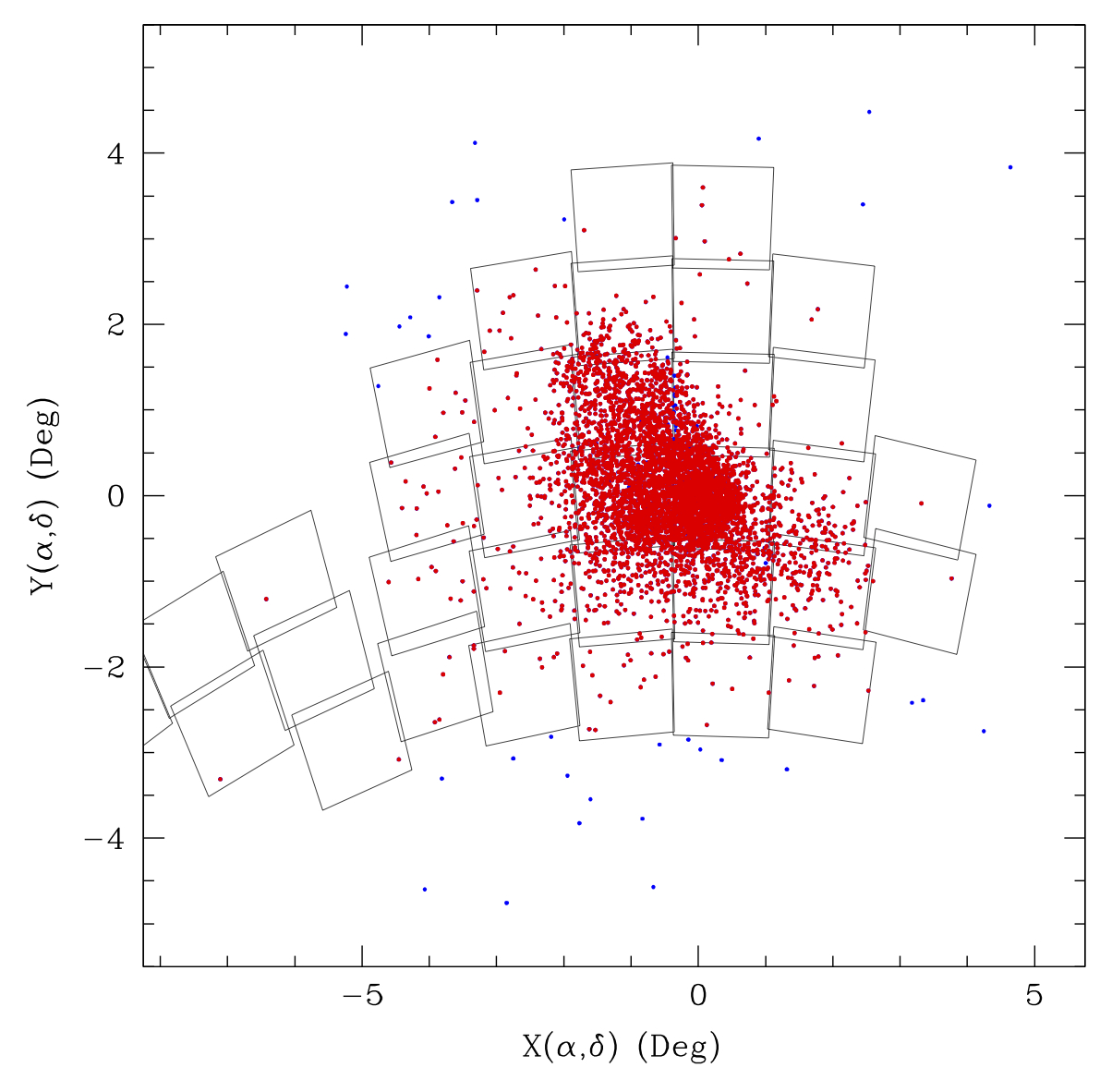}
\caption{Map of the CCs in the SMC. Red and blue circles show CCs in 
  the OGLE\,IV survey with and without counterparts in the VMC survey,
  respectively.  The solid rectangles show the location of the VMC tiles. The centre is at  
  $\alpha_0$=12.54 deg; $\delta_0=-73.11$ deg. 
    \label{fig:map}}
\end{figure}

The scope of the present work is to take advantage of the high precision
NIR $PW$ to estimate accurate single relative distances for each CC in
the SMC, and, in turn, use these data to reveal the 3D structure of the SMC over an area of more
than 40 deg$^{2}$ covering entirely the SMC body and large parts of its
outskirts. The advantage of our data with respect to, e.g. \citet{Scowcroft2016}
is that we have a comparable precision in the photometry but an
incomparably larger sample. As for the work by \citet{Jacy2016}, we
have a comparable sample size, but a higher precision of the single CC
distances. Indeed, going to the NIR has several advantages for the CCs  as
amplitudes are much smaller than in the optical bands, and
hence accurate average magnitudes can be obtained using a smaller number of
epochs along the pulsation cycle with respect to the optical
bands. A particular advantage of the $PW$ using the $(V,K_\mathrm{s})$
bands is that the colour coefficient is the same as that of the $PLC$. This
means that the $PW$ for this particular combination of colours is not 
only reddening-free and mildy metallicity--dependent \citep[see
e.g.][]{Caputo2000}, but also devoid of the uncertainties due to the
finite width of the instability strip affecting the $PL$ relations and
to a lesser extent the $PW$ ones.

\begin{table}
\scriptsize 
\caption{$Y,~J$ and $K_\mathrm{s}$ time--series photometry for the 757 
  CCs investigated in this paper (see text). The sample data below refer to the variable OGLE-SMC-CEP-3294.}
\label{table:VMCPhot}
\begin{center}
\begin{tabular}{ccc}
\hline 
\noalign{\smallskip} 
HJD-2\,400\,000 & $Y$  & $\sigma_{Y}$  \\
\noalign{\smallskip}
\hline 
\noalign{\smallskip} 
   55492.59442  &  16.142   &  0.005  \\
   55492.62957  &  16.154   &  0.005 \\
   55497.70358  &  16.006   &  0.005 \\
   55539.61982  &  15.867   &   0.005 \\
\noalign{\smallskip}
\hline 
\noalign{\smallskip} 
HJD-2\,400\,000 & $J$  & $\sigma_{J}$  \\
\noalign{\smallskip}
\hline 
\noalign{\smallskip} 
   55493.58657  &  15.783   &  0.005  \\
   55493.62563   & 15.792   &  0.005 \\
   55495.55133  &  15.912   &  0.007 \\
   55539.63935  &  15.698   &  0.006 \\
   55778.75250  &  15.811  &   0.006 \\
\noalign{\smallskip}
\hline 
\noalign{\smallskip} 
HJD-2\,400\,000 & $K_\mathrm{s}$  & $\sigma_{K_\mathrm{s}}$  \\
\noalign{\smallskip}
\hline 
\noalign{\smallskip} 
   55493.78578  &  15.592  &   0.010  \\
   55495.57476  &  15.649  &   0.013  \\
   55495.68414  &  15.627  &   0.010  \\
   55497.72311  &  15.553  &   0.011  \\
   55538.61986  &  15.564  &   0.010  \\
   55549.58433  &  15.561  &   0.010  \\
   55769.75246  &  15.650  &   0.011  \\
   55778.77203  &  15.586  &   0.012  \\
   55791.76028  &  15.601  &   0.010  \\
   55818.72858  &  15.640  &   0.011  \\
   55820.67384  &  15.540  &   0.009  \\
   55879.55444  &  15.552  &   0.010  \\
   55880.61689  &  15.541  &   0.011  \\
   55900.56920  &  15.634  &   0.010  \\
   56130.79148  &  15.599  &   0.010  \\
   56173.70146  &  15.579  &   0.011  \\
   56195.63840  &  15.546  &   0.009  \\
   56223.53971  &  15.604  &   0.010  \\
\noalign{\smallskip}
\hline 
\noalign{\smallskip}
\end{tabular}
\end{center}
Table~\ref{table:VMCPhot} is published in its entirety only in the 
electronic edition of the journal. 
A portion is shown here for guidance regarding its form and content. 
\end{table}

This paper is organized as follows: Section 2 presents the
observations and the analysis of the CCs' light curves; Section 3
illustrates the construction of the new $PL$ and $PW$ relationships 
obtained in this paper from the enlarged sample of CCs; 
in Section 4 we find individual distances to CCs in the SMC and discuss
their 3D distribution; Section 5 presents a discussion of the results,
while Section 6 briefly summarizes the main outcomes of this paper.

\section{SMC Classical cepheids in the VMC survey}

The list of CCs in the SMC used as reference was 
taken from the OGLE\,IV survey
\citep[][]{Soszynski2015a,Soszynski2015b,Udalski2015},  whose results
supersede those of the OGLE\,III \citep[][]{Soszynski2010} and EROS\,2
\citep[][]{Tisserand2007} surveys used in Paper I.  In more detail,
OGLE\,IV published the identification, the $V,I$ light curves, and
main properties (periods, mean magnitudes, amplitudes etc.) for 4915
CCs in the SMC\footnote{Note that \citet{Soszynski2015a} estimated the
  OGLE\,IV collection of CCs in the MCs  is complete at least at the 99\% level.}

\begin{table*}
\scriptsize\tiny  
\caption{Photometric results for all the 4793 CCs
  analyzed in this paper. Columns: (1) Identification from OGLE\,IV; (2)  Mode: F=Fundamental; 1O=First
  Overtone; 2O=Second Overtone; 3O=Third Overtone; (3)  $VMC$ tile in which the object is found; (4) Period; (5)
  Number of epochs in $Y$; (6)--(7) Intensity-averaged magnitude in
  $Y$ and relative uncertainty; (8)--(9) Peak-to-peak amplitude in
  $Y$ and relative uncertainty; (10) to (14) As for column (5) to (9)
  but for the $J$ band; (15) to (19) As for column (5) to (9) but for
  the $K_\mathrm{s}$ band; (20) $E(V-I)$ values adopted in this work.}
\label{table:averages}
\begin{center}
\begin{tabular}{cccccccccccccccccccc}
\hline  
\noalign{\smallskip}   
       OGLE\_ID    & MODE & VMC TILE &     P     &  ${\rm n}_Y$ & $\langle Y \rangle$ & $\sigma_{\langle Y \rangle}$&   A($Y$) &  $\sigma_{{\rm A}(Y)}$ &  ${\rm n}_J$ & $\langle J \rangle$ & $\sigma_{\langle J \rangle}$&   A($J$) &  $\sigma_{{\rm A}(J)}$  &  ${\rm n}_{K_\mathrm{s}}$ & $\langle K_\mathrm{s} \rangle$ & $\sigma_{\langle K_\mathrm{s} \rangle}$&   A($K_\mathrm{s}$) &  $\sigma_{{\rm A}(K_\mathrm{s})}$ & E(V-I) \\
   & & & d & & mag & mag &mag & mag & & mag & mag &mag & mag& & mag &   mag &mag & mag & mag \\                     
(1)    & (2)  & (3) & (4) & (5) &(6)  & (7) &(8) & (9) & (10) & (11) & (12)  & (13) & (14) & (15) &(16)  & (17) &(18) & (19) & (20)  \\                     
\noalign{\smallskip}
\hline  
\noalign{\smallskip}    
   OGLE-SMC-CEP-2476  &        1O  &   SMC\_5\_4   &    0.2526028  &    4  &   19.055  &    0.072  &   0.11  &   0.09  &    5  &   18.875  &    0.033  &   0.16  &   0.07  &   18  &   18.696  &    0.046 &   0.120 &   0.054  &   0.05  \\  
   OGLE-SMC-CEP-3867  &  1O/2O/3O  &   SMC\_4\_4   &    0.2688496  &    4  &   18.454  &    0.022  &   0.19  &   0.03  &    5  &   18.247  &    0.019  &   0.14  &   0.03  &   15  &   18.033  &    0.033 &   0.050 &   0.048  &   0.05  \\  
   OGLE-SMC-CEP-2507  &     1O/2O  &   SMC\_4\_4   &    0.2775568  &    4  &   18.651  &    0.017  &   0.07  &   0.04  &    5  &   18.428  &    0.022  &   0.07  &   0.04  &   15  &   18.206  &    0.041 &   0.197 &   0.084  &   0.03  \\  
   OGLE-SMC-CEP-2752  &        1O  &   SMC\_4\_4   &    0.2837457  &    8  &   17.954  &    0.011  &   0.12  &   0.03  &   10  &   17.809  &    0.011  &   0.15  &   0.02  &   33  &   17.683  &    0.016 &   0.100 &   0.026  &   0.11  \\  
   OGLE-SMC-CEP-2095  &        1O  &   SMC\_5\_3   &    0.2870474  &    5  &   18.129  &    0.016  &   0.33  &   0.02  &    9  &   17.918  &    0.009  &   0.18  &   0.04  &   19  &   17.774  &    0.019 &   0.064 &   0.039  &   0.03  \\  
   OGLE-SMC-CEP-0022  &        1O  &   SMC\_4\_2   &    0.3136645  &    5  &   18.445  &    0.019  &   0.18  &   0.03  &    5  &   18.268  &    0.016  &   0.10  &   0.02  &   15  &   18.057  &    0.033 &   0.099 &   0.053  &   0.02  \\  
   OGLE-SMC-CEP-4548  &        1O  &   SMC\_5\_5   &    0.3247217  &    5  &   18.402  &    0.019  &   0.13  &   0.03  &    4  &   18.245  &    0.027  &   0.12  &   0.02  &   14  &   18.164  &    0.029 &   0.133 &   0.054  &   0.03  \\  
   OGLE-SMC-CEP-1471  &     1O/2O  &   SMC\_5\_3   &    0.3271798  &    5  &   17.726  &    0.011  &   0.11  &   0.01  &    9  &   17.544  &    0.007  &   0.11  &   0.02  &   19  &   17.327  &    0.014 &   0.062 &   0.021  &   0.03  \\  
   OGLE-SMC-CEP-2527  &     1O/2O  &   SMC\_4\_4   &    0.3292266  &    4  &   18.173  &    0.024  &   0.07  &   0.03  &    5  &   17.941  &    0.015  &   0.06  &   0.02  &   15  &   17.676  &    0.027 &   0.100 &   0.046  &   0.03  \\  
   OGLE-SMC-CEP-2683  &        1O  &   SMC\_4\_4   &    0.3373664  &    4  &   18.045  &    0.011  &   0.10  &   0.02  &    5  &   17.751  &    0.013  &   0.02  &   0.02  &   15  &   17.350  &    0.021 &   0.070 &   0.007  &   0.05  \\  
   OGLE-SMC-CEP-3287  &        1O  &   SMC\_4\_4   &    0.3464235  &    4  &   18.061  &    0.017  &   0.14  &   0.04  &    5  &   17.876  &    0.016  &   0.11  &   0.03  &   15  &   17.677  &    0.023 &   0.131 &   0.035  &   0.04  \\  
   OGLE-SMC-CEP-4242  &        1O  &   SMC\_4\_4   &    0.3465840  &    4  &   17.859  &    0.008  &   0.06  &   0.02  &    5  &   17.563  &    0.010  &   0.06  &   0.02  &   15  &   17.157  &    0.013 &   0.076 &   0.026  &   0.05  \\  
   OGLE-SMC-CEP-1606  &     1O/2O  &   SMC\_4\_3   &    0.3525089  &    6  &   18.310  &    0.018  &   0.20  &   0.04  &    6  &   18.073  &    0.019  &   0.12  &   0.04  &   16  &   17.784  &    0.027 &   0.070 &   0.030  &   0.02  \\  
   OGLE-SMC-CEP-4628  &        1O  &   SMC\_3\_5   &    0.3643637  &    7  &   17.390  &    0.006  &   0.05  &   0.01  &    5  &   17.106  &    0.007  &   0.02  &   0.01  &   16  &   16.642  &    0.009 &   0.016 &   0.011  &   0.05  \\  
   OGLE-SMC-CEP-3784  &        1O  &   SMC\_4\_4   &    0.3803699  &    4  &   18.275  &    0.047  &   0.18  &   0.02  &    5  &   18.007  &    0.015  &   0.19  &   0.03  &   15  &   17.745  &    0.025 &   0.090 &   0.036  &   0.03  \\  
   OGLE-SMC-CEP-3660  &     1O/2O  &   SMC\_4\_4   &    0.3867429  &    4  &   16.895  &    0.008  &   0.12  &   0.01  &    5  &   16.539  &    0.006  &   0.08  &   0.01  &   15  &   16.136  &    0.006 &   0.034 &   0.013  &   0.05  \\  
   OGLE-SMC-CEP-4243  &        1O  &   SMC\_4\_4   &    0.3937253  &    4  &   18.126  &    0.012  &   0.28  &   0.03  &    5  &   17.909  &    0.012  &   0.14  &   0.03  &   15  &   17.642  &    0.019 &   0.068 &   0.036  &   0.09  \\  
   OGLE-SMC-CEP-0310  &        1O  &   SMC\_4\_3   &    0.3942412  &    6  &   17.896  &    0.015  &   0.19  &   0.06  &    6  &   17.703  &    0.012  &   0.16  &   0.03  &   16  &   17.531  &    0.019 &   0.150 &   0.033  &   0.06  \\  
   OGLE-SMC-CEP-1357  &        2O  &   SMC\_4\_3   &    0.4012875  &    6  &   17.701  &    0.010  &   0.04  &   0.02  &    6  &   17.461  &    0.011  &   0.03  &   0.02  &   16  &   17.256  &    0.015 &   0.043 &   0.025  &   0.05  \\  
   OGLE-SMC-CEP-2265  &        1O  &   SMC\_3\_3   &    0.4085550  &    7  &   17.824  &    0.015  &   0.13  &   0.03  &    5  &   17.688  &    0.017  &   0.21  &   0.02  &   18  &   17.500  &    0.017 &   0.100 &   0.032  &   0.03  \\  
\noalign{\smallskip}
\hline  
\noalign{\smallskip}
\end{tabular}
Table \ref{table:averages} is published in its entirety in the
electronic edition of the MNRAS.  A portion is
shown here for guidance regarding its form and content.
\end{center}
\end{table*}

\begin{figure}
\includegraphics[width=8.5cm]{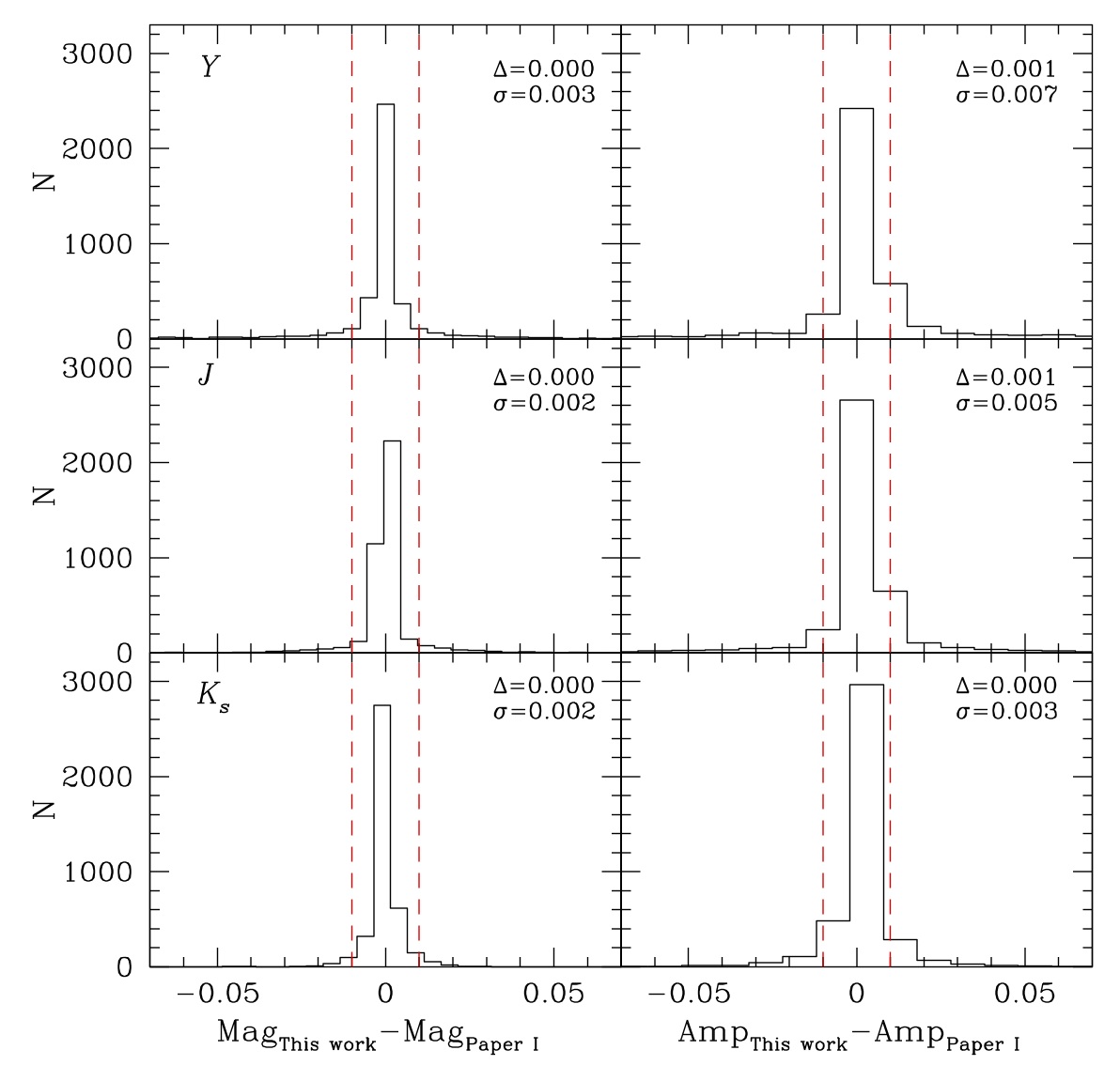}
\caption{Left panels: comparison between the intensity-averaged 
  magnitudes obtained in this work and those of Paper I. Right panels: same as Left but for the amplitudes. In 
  all panels the dashed lines show the $\pm$0.01 mag limits. The bins
  widths   are 0.005 and 0.01 mag for the magnitudes and amplitudes,
  respectively. Each panel shows the average difference ($\Delta$) and
  the relative dispersion ($\sigma$), calculated with a 3-$\sigma$
  clipping procedure. 
    \label{fig:compPapI}}
\end{figure}

\begin{figure}
\includegraphics[width=8.5cm]{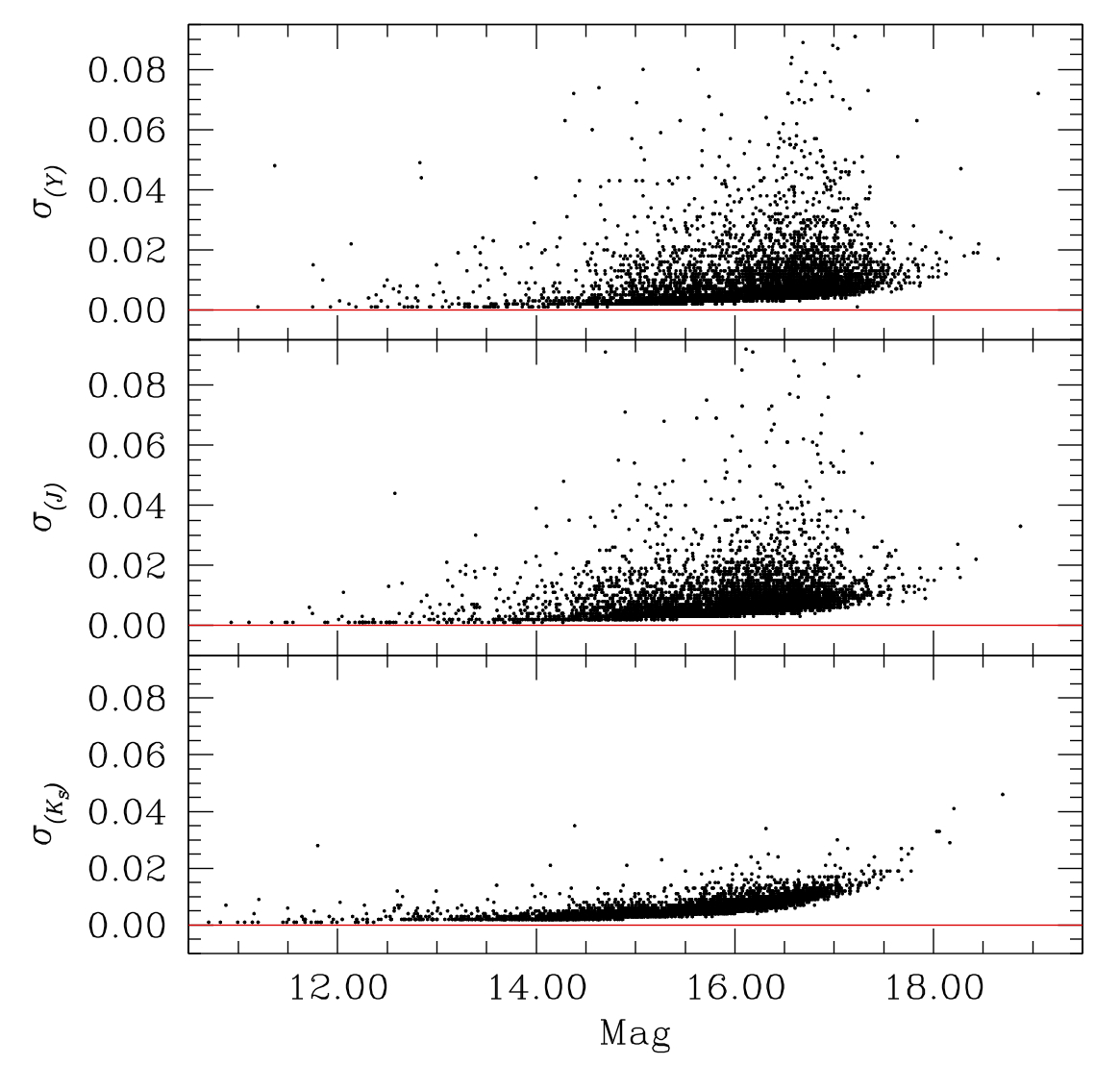}
\caption{Uncertainties in our photometry according to the Monte Carlo simulations.  
    \label{fig:magAcc}}
\end{figure}

 The area observed by the VMC survey in the region of the SMC is shown
in Fig.~\ref{fig:map}, where each square represents one VMC tile (each
tile is 1.65 deg$^2$ on the sky).  In this paper we present results for
the CCs included in 31 tiles completely or nearly completely observed,
processed and catalogued by the VMC survey as of 2016 August 22
(including observations until 2016 March 31). The cross-match between
the OGLE\,IV CCs and VMC sources with at least six epochs in
$K_\mathrm{s}$ (sufficient to obtain very precise magnitudes) 
within a radius of 0.5 arcsec returned 4784 matches. Of
the 131 missing objects, 55 are located outside of the VMC tiles or in the gap
between tiles SMC\,5\_3 and 5\_4\footnote{Note that this gap will be
  filled with ad hoc observations already approved by ESO;
  proposal 099.D-0194.}, whereas for the remaining 76 we enlarged the
search radius to 2 arcsec and inspected the light curve of the
closest matching star, as well as the actual position of the targets
on the VMC images. As a result, we recovered nine additional CCs with
usable light curves, while the remaining 67 objects showed unusable
data due to different causes, such as blending, too few epochs
(vicinity to borders of the tile) and saturation. After this
procedure, the total number of stars with usable light curves contains
4793 objects, which represent the 97.5\% of the OGLE\,IV sample. The
classification of the targets in terms of pulsation modes was also
taken from OGLE\,IV. Our sample counts 2684, 1729, 90 and 290
F, 1O, 2O, and mixed
modes (F/1O, 1O/2O, F/1O/2O, 1O/2O/3O, where TO stands for Third
Overtone) CCs, respectively.  Note that with respect
to Paper I, the sample analysed here is significantly larger
($\sim$720 additional CCs) and more accurate because: i) OGLE\,IV covers a larger
area with respect to OGLE\,III and EROS\,2; ii) OGLE\,IV reclassified
with different variable types 32 and 10 objects originally classified
as CCs in OGLE\,III and
EROS\,2, respectively. In total the present sample has 4076 stars in
common with that of Paper I.

A general description of the observations in the context of the VMC
survey can be found in \citet{Cioni2011}, whereas the procedures
adopted to study the variable stars were discussed in detail in Paper
I and in \citet{Ripepi2012a,Ripepi2012b,Moretti2014,Muraveva2014,
Ripepi2014b,Muraveva2015,Ripepi2015,Moretti2016,Marconi2017}.
Therefore, here we only briefly recall that the VMC
$K_\mathrm{s}$-band time--series observations were programmed to span
13 separate epochs executed over several consecutive months. This
observing strategy allowed us to obtain well-sampled light curves for
the relevant pulsating stars.  As for the $Y$ and $J$ bands, the planned number of
epochs is four (two of these epochs are obtained with half exposure
times).  However, additional observations (a few epochs) are usually 
available for each tile (in particular for the $K_\mathrm{s}$-band)
because some Observing Blocks (OBs) executed out of specifications provided usable
data. Additionally, the overlap between the tiles and the high
density of CCs in the body of the SMC cause the presence of many CCs
in two or more different tiles and consequently about 490 CC light
curves are sampled with more than 23 epochs.  The situation is similar
in the $Y$ and $J$ bands, and the number of CCs with more than 10
epochs is 232 and 404 in the $Y$ and $J$ filters, respectively (see
Paper I for more details).  Again, there are some differences with
respect to Paper I, as additional VMC observations were obtained and
fully reduced since the time of the writing of Paper I. As a result,
considering the sample in common with Paper I, we have now 40
pulsators with approximately twice the number of planned epochs in the
$YJK_\mathrm{s}$--bands. In
addition, 1229 stars have one epoch more in $K_\mathrm{s}$, while 243
objects show between 2 and 5 additional epochs again in the
$K_\mathrm{s}$. As we shall see in what follows, the availability of
additional epochs of observations for a subsample of pulsators can
lead to slight differences in the measurements of their magnitudes.

The data analysed in this paper were processed by means of the pipeline of the
VISTA Data Flow System \citep[VDFS,][]{Emerson2004,Irwin2004}. The photometry
is in the VISTA photometric system \citep[Vegamag=0; the VISTA
photometric system is described by][]{Carlos2017}. The time--series data
used in this work were downloaded from the VISTA Science
Archive\footnote{http://horus.roe.ac.uk/vsa/}
\citep[VSA,][]{Cross2012}. More details about the characteristics of
CC  time--series such as those used in this paper can be found in Paper I.

The VMC light curves in the $Y,~J$ and $K_\mathrm{s}$  filters of  the 717 CCs analysed in this paper and not
 present in Paper I are shown in Appendix~\ref{appendixA}. 
Similarly,  in  Table~\ref{table:VMCPhot} we publish the VMC photometry for these
757 variables (including the quoted 40 variables having almost twice
epochs with respect to Paper I). The complete versions of the table and of the figures are available online on the journal's site.

\section{Average magnitudes and $PL/PW$ relations}

We estimated the $Y,~J$ and $K_\mathrm{s}$  intensity-averaged magnitudes and the peak-to-peak amplitudes 
for the full sample of 4793 CCs adopting the same techniques as 
in Paper I. In brief, we constructed eight different templates for each
of the three $Y,~J$ and $K_\mathrm{s}$ bands and used a modified
$\chi^2$ technique to identify the best-fitting template. To estimate
the uncertainties on the mean magnitudes and the peak-to-peak  amplitudes,
we adopted a Monte Carlo technique consisting of the generation of 100
mock time--series for each star and each filter, to which Gaussian noise 
was added with standard deviations $\sigma$s corresponding to the average errors on the phase points
for the target star. At variance with Paper I, we decided to use the
Monte Carlo simulations to estimate not only the errors on the
amplitudes in the three bands, but also the actual amplitudes, by
 averaging  the amplitudes resulting from the 100 experiments. 

We did not use the simulation to estimate the intensity-averaged magnitudes
 (but only the uncertainties on their values) because we verified that
 the difference between the magnitudes obtained from the best-fit and
 those estimated as the average of the 100 simulations, agree very well
 within the uncertainties. This is because the magnitudes estimated with the
 templates are only mildly affected  by even large errors in the estimated
 amplitudes, an occurrence that can happen when we have only 4-6
 epochs in the light curves (e.g. in the $Y\,J$ bands). 

The result of these procedures is shown in
Table~\ref{table:averages} which is available in its entirety
associated with the
online version of the paper. 

Figure~\ref{fig:compPapI} shows the difference between the present  
and Paper I photometry for the 4076 stars in common. It can be seen  
that for the vast majority of the stars the difference in magnitude  
and amplitude is well below 0.01 mag.  
Figure~\ref{fig:magAcc} gives an  
idea of the typical errors as a function of the magnitudes for the  
whole sample of 4793 CCs.

The intensity--averaged magnitudes thus obtained were used to construct 
new $PL$ and $PW$ relations including the new CCs. These relations will be at the basis of our
structural analysis of the SMC. 
Before proceeding, we have to calculate the dereddened $Y,~J$ and
$K_\mathrm{s}$ magnitudes. The last column of Table~\ref{table:averages}  reports
the individual colour excesses $E(V-I)$ derived according to
\citet{Haschke2011}. For the CCs located outside the region investigated by
those authors (i.e. the area covered by the OGLE\,III survey), we
adopted a mean value of $E(V-I)$=0.035 mag, calculated averaging the
$E(V-I)$ values obtained for the CCs inside the area covered by OGLE\,III. For the extinction
corrections, as in Paper I, we used \citet{Cardelli1989,Kerber2009,Gao2013}.

\begin{figure}
\includegraphics[width=8.5cm]{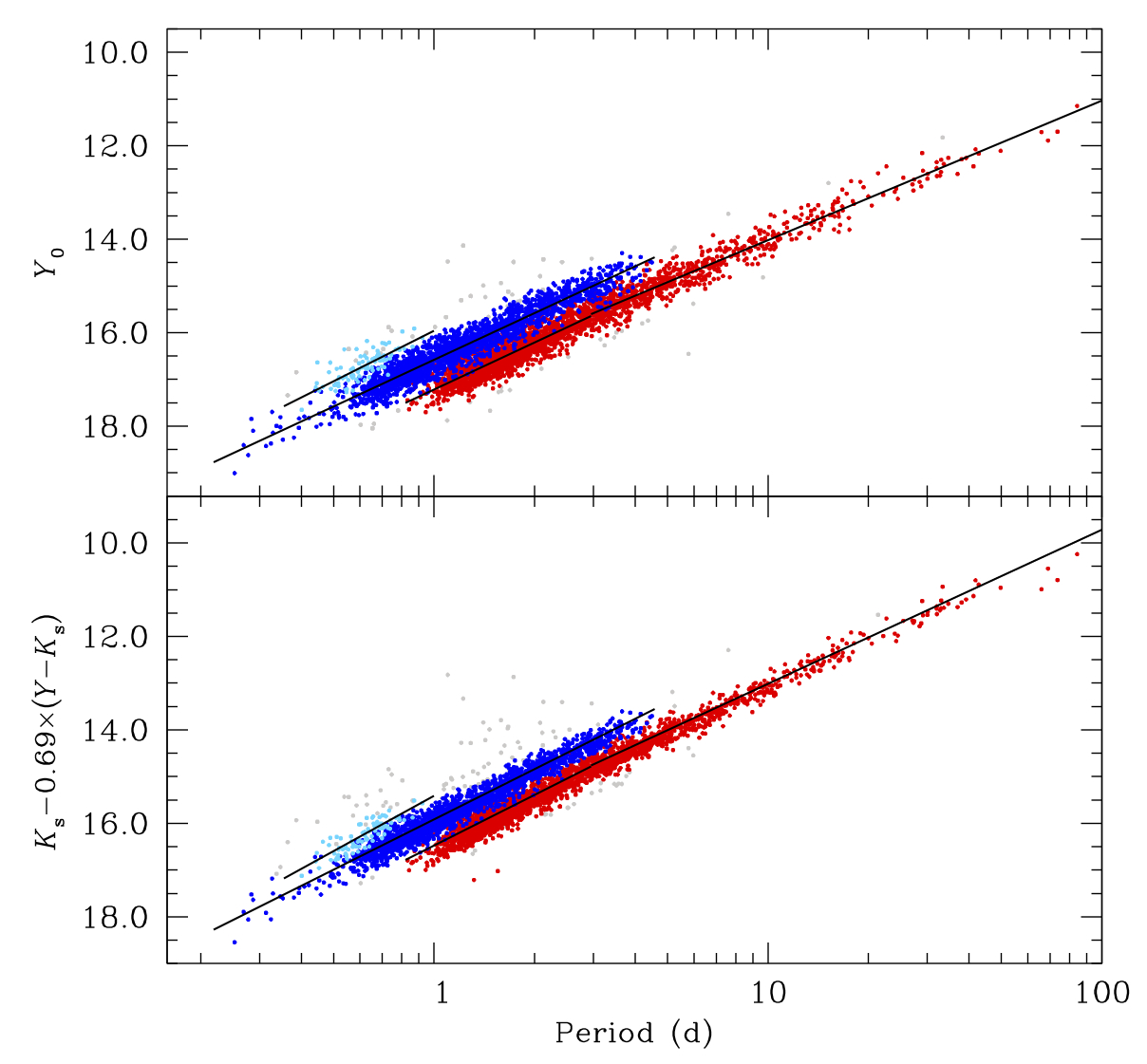}
\caption{$PL(Y)$ and $PW(Ks, Y)$ relations for 
the SMC CCs investigated in this paper. F, 1O and 2O
pulsators are shown as red, blue and light blue filled circles, respectively. The
grey filled circles show objects excluded from the analysis because they 
deviate more than 3.5$\sigma$ from the best-fits to the data
represented by solid lines.
    \label{fig:figY}}
\end{figure}

\begin{figure}
\includegraphics[width=8.5cm]{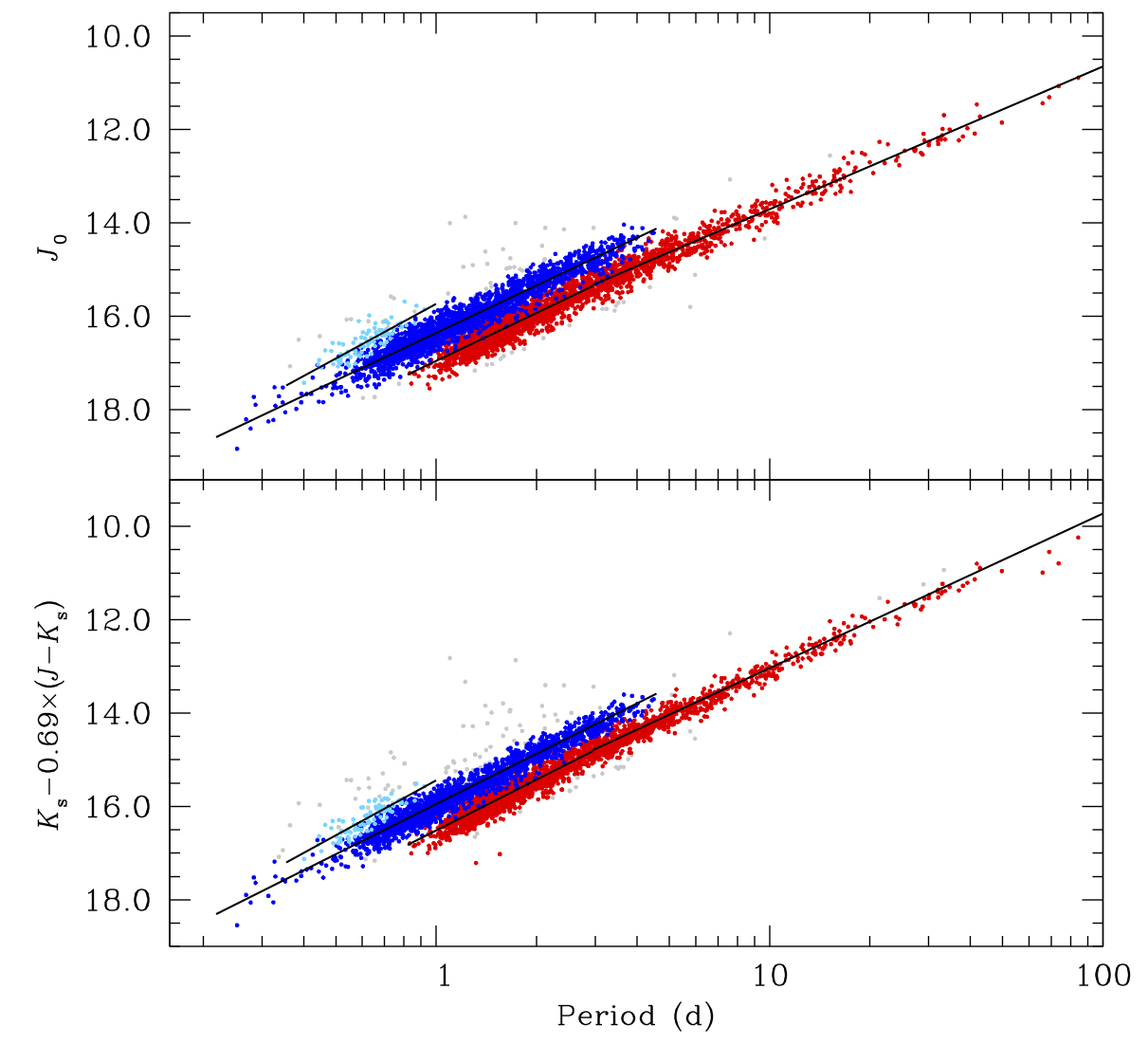}
\caption{As in Fig.~\ref{fig:figY}, but for the $J$-filter.
    \label{fig:figJ}}
\end{figure}

\begin{figure}
\includegraphics[width=8.5cm]{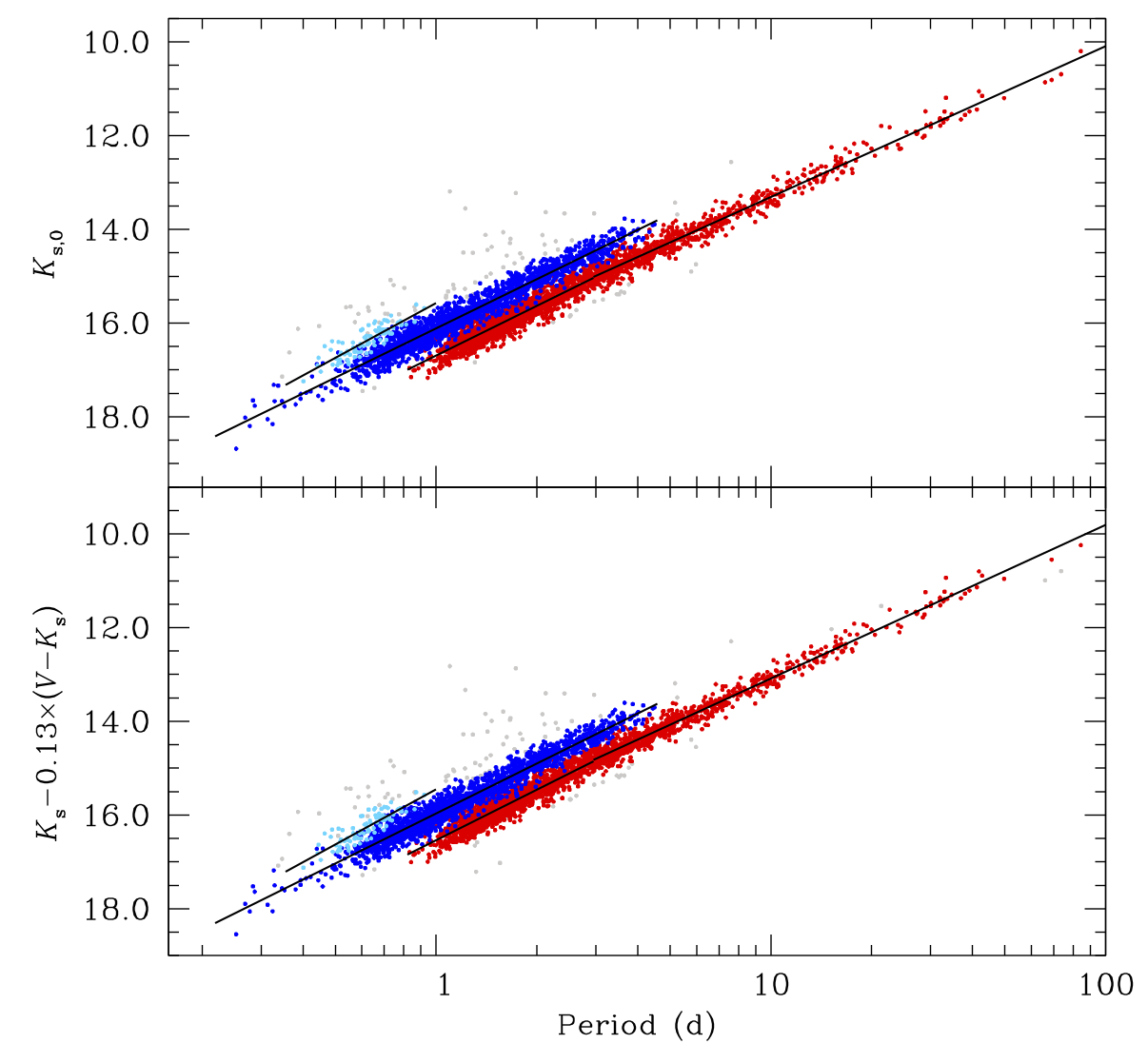}
\caption{As in Fig.~\ref{fig:figY}, but for the $V$-filter.
    \label{fig:figK}}
\end{figure}

As discussed in detail in Paper I, we
fitted two relations to the F-mode pulsators because of the presence
of a change in slope at Period$\sim$2.95 d. 
We chose not to recalculate the $PLC$ relations
because they are substantially very similar to the $PW$ relations
(almost identical in the case of the $PW(V,K_\mathrm{s})$). With
respect to Paper I we used a slightly different technique to perform
the least--squares fit to the data. Specifically, we applied to all
relations a sigma-clipping algorithm, fixing 3.5$\sigma$. This
approach guarantees repeatability of the procedure and avoids
excluding stars that  could be off the
ridge-line due to their position and not to photometric problems. 
The results of these procedures are shown in
Figs.~\ref{fig:figY},~\ref{fig:figJ} and \ref{fig:figK} and
 in Table~\ref{table:pl}. A comparison between Table~\ref{table:pl}
 and Table 6 of Paper I shows that the parameters of the 
$PL$ and $PW$ relations (and relative uncertainties) agree within the
errors.

\begin{table}
\scriptsize
\caption{$PL$ and $PW$ relations for F, 1O and 2O CCs. The
Wesenheit functions are defined in the table. \label{table:pl}}
\begin{center}
\begin{tabular}{lccccc}
\hline 
\noalign{\smallskip} 
Mode    &   $a$ & $\sigma a$&   $b$ & $\sigma b$ & rms \\
\noalign{\smallskip} 
\noalign{\smallskip} 
\hline 
\noalign{\smallskip} 
\multicolumn{6}{c}{$Y^0$=$a$ log$P$+$b$}\\
\noalign{\smallskip} 
\hline 
\noalign{\smallskip}  
F log$P<0.47$     & -3.353   &  0.041  &  17.216 &    0.010   &  0.202   \\
F log$P\ge0.47$   & -2.987   &  0.027  &  17.008 &    0.021   &  0.202   \\
1O                & -3.321   &  0.023  &  16.582 &    0.005   &  0.215   \\
2O                & -3.572   &  0.305  &  15.964 &    0.068   &  0.205   \\
\noalign{\smallskip} 
\hline 
\noalign{\smallskip} 
\multicolumn{6}{c}{$J^0$=$a$ log$P$+$b$}	      \\		                         
\noalign{\smallskip} 
\hline 
\noalign{\smallskip} 
F log$P<0.47$     & -3.406   &  0.041  &  16.952 &    0.010   &  0.206   \\
F log$P\ge0.47$   & -3.070   &  0.026  &  16.778 &    0.021   &  0.196   \\
1O                & -3.377   &  0.021  &  16.356 &    0.005   &  0.203   \\
2O                & -3.602   &  0.308  &  15.776 &    0.069   &  0.207   \\
\noalign{\smallskip} 
\hline 
\noalign{\smallskip} 
\multicolumn{6}{c}{$K_\mathrm{s}^0$=$a$ log$P$+$b$}\\
\noalign{\smallskip} 
\hline 
\noalign{\smallskip} 
F log$P<0.47$     & -3.513   &  0.036  &  16.686 &    0.009   &  0.177   \\
F log$P\ge0.47$   & -3.224   &  0.023  &  16.530 &    0.018   &  0.167   \\
1O                & -3.489   &  0.020  &  16.112 &    0.005   &  0.186    \\
2O                & -3.989   &  0.317  &  15.520 &    0.070   &  0.208    \\
\noalign{\smallskip} 
\hline 
\noalign{\smallskip} 
\multicolumn{6}{c}{$W(Y,K_\mathrm{s})$=$K_\mathrm{s}-0.42\,(Y-K_\mathrm{s})$=$a$ log$P$+$b$}\\
\noalign{\smallskip} 
\hline 
\noalign{\smallskip} 
F log$P<0.47$     & -3.608   &  0.033  &  16.473 &    0.008   &  0.165   \\
F log$P\ge0.47$   & -3.296   &  0.020  &  16.308 &    0.016   &  0.149   \\
1O                & -3.577   &  0.018  &  15.918 &    0.004   &  0.173   \\
2O                & -3.914   &  0.271  &  15.415 &    0.060   &  0.174   \\
\noalign{\smallskip} 
\hline 
\noalign{\smallskip} 
\multicolumn{6}{c}{$W(J,K_\mathrm{s})$=$K_\mathrm{s}-0.69\,(J-K_\mathrm{s})$=$a$ log$P$+$b$}\\
\noalign{\smallskip} 
\hline 
\noalign{\smallskip} 
F log$P<0.47$     & -3.597   &  0.034  &  16.504 &    0.009   &  0.169   \\
F log$P\ge0.47$   & -3.334   &  0.021  &  16.363 &    0.017   &  0.158   \\
1O                & -3.572   &  0.019  &  15.945 &    0.004   &  0.179   \\
2O                & -4.060   &  0.353  &  15.374 &    0.078   &  0.232   \\
\noalign{\smallskip} 
\hline 
\noalign{\smallskip} 
\multicolumn{6}{c}{$W(V,K_\mathrm{s})$=$K_\mathrm{s}-0.13\,(V-K_\mathrm{s})$=$a$ log$P$+$b$}\\
\noalign{\smallskip} 
\hline 
\noalign{\smallskip} 
F log$P<0.47$     & -3.567   &  0.034  &  16.527 &    0.009   &  0.170   \\
F log$P\ge0.47$   & -3.291   &  0.021  &  16.375 &    0.017   &  0.155   \\
1O                & -3.544   &  0.019  &  15.967 &    0.004   &  0.178   \\
2O                & -3.790   &  0.344  &  15.438 &    0.077   &  0.231   \\
\noalign{\smallskip}
\hline
\end{tabular}
\end{center}
\end{table}

\begin{figure}
\includegraphics[width=8.5cm]{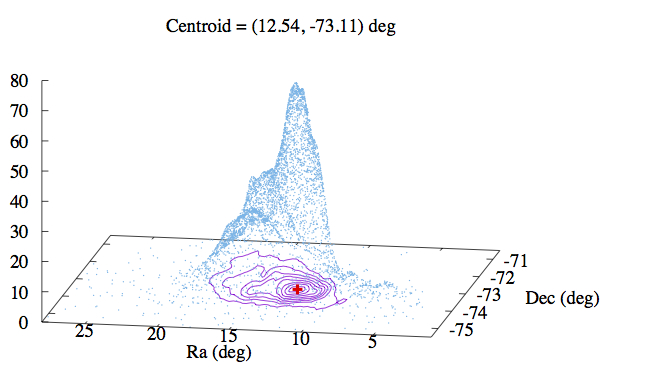}
\caption{
Bi-dimensional histogram of the spatial distribution in $\alpha,\delta$ of CCs 
in the SMC. The contour levels span the range 0-40 counts with an 
interval of 5 counts between each level. A cross marks the chosen centre of 
the distribution. 
    \label{fig:centre}}
\end{figure}

\section{The 3D structure of the SMC young population}

The scope of this section is to determine the three--dimensional structure
of the young population of the SMC as traced by the CCs. 
The first step is to estimate the relative distances of each CC with respect to
 the centre of the distribution of the CCs in the SMC \citep[note that the centre of the SMC is poorly defined, see
discussion in][]{deGrijs2015}. 
To this aim we produced a smoothed 2D histogram in the equatorial
coordinates 
$\alpha,\delta$ of the studied variables using a box-smoothing area of
0.20$\times$0.20 deg$^2$ (but the result is only mildly dependent on this
choice), as shown in Fig.~\ref{fig:centre}. This figure reveals that the spatial distribution
of the CCs is rather asymmetric and to determine its centre we
calculated 
the average $\alpha,\delta$ using only the bi-dimensional bins
containing more than 60 counts (we verified that the result is not significantly
affected by this choice). The resulting centre is
$\alpha_0$=12.54$\pm$0.01 deg and $\delta_0=-73.11\pm$0.01 deg, the
uncertainties are the rms of the mean. 
This result is compared with the literature in
Table~\ref{table:centre}. Our result is compatible with those estimated from the density of 
the stars \citep{Gonidakis2009,Rubele2015} and consistent in $\delta$ but
differing by about one degree in $\alpha$ from the previous values inferred from the
distribution of  RR Lyrae and CCs variables 
\citep[][respectively]{Subramanian2012,Subramanian2015}.

The next step is to find the relative distances of each CC with
respect to this centre. To this aim we decided to use only the $PW$
relations derived in the previous section, because: i) they are
intrinsically more accurate with respect to the $PL$ relations; ii)
they are independent of reddening; iii) especially in the case of
$W(V,K_\mathrm{s})$ the colour term is very similar to that of the
$PLC$ relations \citep[see e.g.][and Paper I]{Ripepi2012b}, which
 should, in principle, show a negligible dispersion \citep[see
e.g.][]{Bono1999}. 

\begin{table*}
\scriptsize 
\caption{Comparison between the centre of the distribution of the SMC 
  CCs found here and a compilation of literature centres for the
  SMC \citep[see also][for a discussion about the definition of the centre of the SMC]{deGrijs2015}.}
\label{table:centre}
\begin{center}
\begin{tabular}{llcc}
\hline 
\noalign{\smallskip} 
Ra$_0$& Dec$_0$ &  Method  & Source  \\
\noalign{\smallskip}
\hline 
\noalign{\smallskip} 
 16.25 & $-$72.42 & Kinematics of \ion{H}{i}   &  \citet{Stanimirovic2004} \\
 12.75 & $-$73.1 & Density of 2MASS K and M stars & \citet{Gonidakis2009} \\
 13.38 & $-$73.0 & Density of RR Lyrae stars& \citet{Subramanian2012} \\
 12.60 & $-$73.09  & Stellar density VMC &   \citet{Rubele2015}  \\
 13.90 & $-$72.98   & Density of CCs (OGLE\,III) &  \citet{Subramanian2015}\\
 12.54 & $-$73.11   &  Density of CCs (OGLE\,IV) & this work \\ 
\noalign{\smallskip}
\hline 
\noalign{\smallskip}
\end{tabular}
\end{center}
\end{table*}

We first calculated the magnitude difference between all CC and the $PW$
relations: 
\begin{equation}
\Delta W(\lambda_1,\lambda_2)_i  =  W(\lambda_1,\lambda_2)_i-(\alpha   + \beta {\rm log}P)
\end{equation}
\noindent 
where $(\lambda_1,\lambda_2)$ can assume the value of
$(V,K_\mathrm{s})$ or $(J,K_\mathrm{s})$ and the relative $\alpha$ and
$\beta$ values are listed in Table~\ref{table:pl}. To transform these
$\Delta W$ values into absolute distances, we adopted the distance modulus of
the SMC we estimated in Paper I, distance modulus $\mu=19.01$ mag 
corresponding to $D_{\rm SMC}=63$ kpc. The precise
value of this quantity does not affect the subsequent analysis
\citep[see e.g.][]{Subramanian2015}. The individual distances are
hence calculated as:

\begin{equation}
D_i = D_{\rm SMC} 10^{(\Delta W_i/5)}
\label{eqn:distance}
\end{equation}

\noindent
As for the errors on the individual relative distances, the photometric errors
 are shown in Fig.~\ref{fig:magAcc}. For the large majority of the
 CCs, they 
 are less than 0.02 mag in $Y$ and $J$, and less than 0.01 mag in
 $K_\mathrm{s}$. This translates at the distance of the SMC in
 $\sim$0.6 and 0.3 kpc, respectively. When using the $PW$ relations the
 photometric errors are summed in quadrature and become of the order 
of 0.65 kpc for all combinations of colours used in this work \citep[we
adopted an uncertainty of 0.02 mag for the OGLE\,IV  Johnson--$V$ band,
see][]{Jacy2016}. The possible
inaccuracy of the reddening law chosen to construct the $PW$ relations could
in principle be an additional cause of uncertainty, but the
coefficients are very small in the NIR and this contribution is hence 
negligible. However, the intrinsic dispersion of the
various relations adopted can represent the major source of error. Any
$PL$, $PW$ or $PLC$ has an intrinsic dispersion, due to e.g. the finite width of
the instability strip, mass--loss, rotation, differences in
metallicity. The major contributor to the intrinsic dispersion is
certainly the finite width of the instability strip, which affects
significantly the $PL$ (even though in NIR bands the effect is
reduced). This effect is much reduced for $PW$ relations. In our case, the most accurate 
relations at our disposal are the $PW(J,K_\mathrm{s})$ and
$PW(V,K_\mathrm{s})$. An estimate of the intrinsic dispersion in  these 
relations is $\sim$0.04 and 0.05 mag, respectively, as reported by 
\citet{Inno2016}, on the basis of the up--to--date
theoretical scenario of \citet[][]{Bono1999,Bono2010}. Hence, summing
up in quadrature all the sources of errors, our 
individual relative distance errors are $\sim$1.35 and 1.6 kpc (at the distance
of the centroid of the SMC) for the $PW(J,K_\mathrm{s})$ and
$PW(V,K_\mathrm{s})$, respectively.

\begin{figure}
\includegraphics[width=8.5cm]{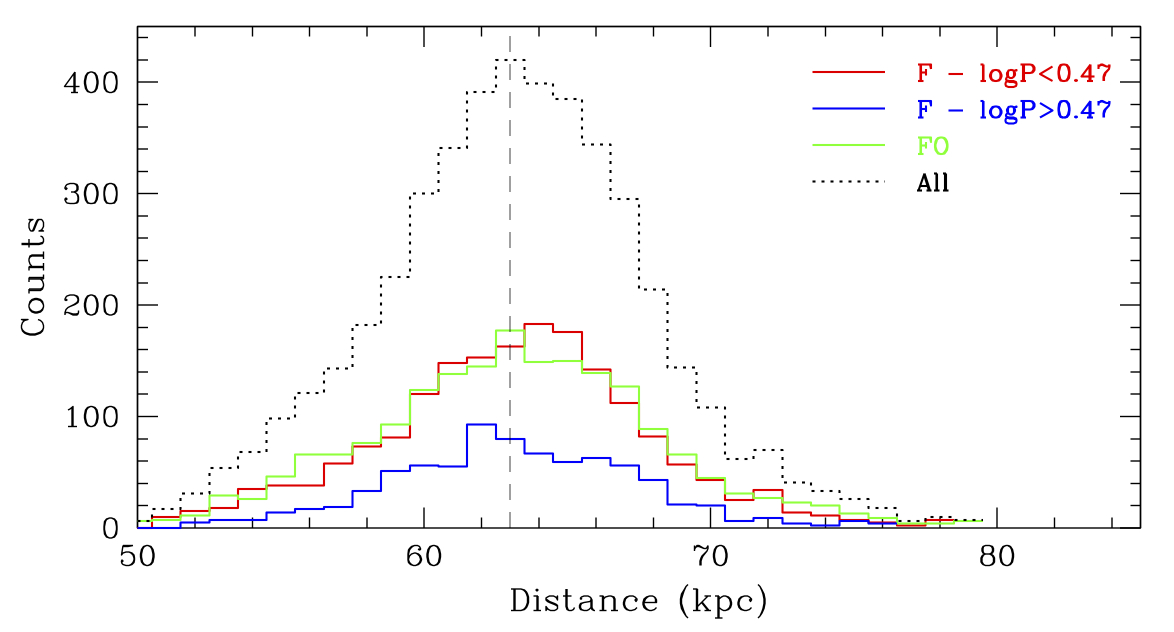}
\caption{Distribution of the individual distances for the CCs 
analysed in this work as obtained by means of the  $W(V,K_\mathrm{s})$
magnitude (see text). The vertical dashed line shows the distance of
the centroid of the SMC. 
    \label{fig:histoDistances}}
\end{figure}

Figure~\ref{fig:histoDistances}
displays the histogram of the results obtained with the
$W(V,K_\mathrm{s})$ (Eq.~\ref{eqn:distance}; very similar results are obtained with
$PW(J,K_\mathrm{s})$). It shows that there is
no systematic difference between the distance distribution of F 
pulsators before/after the break and 1O pulsators. 

Having estimated the individual distances we are now able to
calculate the cartesian coordinates for the CCs investigated here. 
According to \citet{vandermarel2001,Weinberg2001} for each CC we can write:

\begin{eqnarray}
X_i &=& -D_i\sin(\alpha_i - \alpha_0)\cos\delta_i  \nonumber \\
Y_i &= &D_i\sin\delta_i\cos\delta_0 - D\sin\delta_0\cos(\alpha_i -
\alpha_0)\cos\delta_i  \nonumber \\
Z_i &= &D\sin\delta_i\sin\delta_0 - D\cos\delta_0\cos(\alpha_i -
\alpha_0)\cos\delta_i , \nonumber 
\label{8}
\end{eqnarray}

\noindent 
where $D_i$ is the distance to each CC calculated based on 
Eq.~\ref{eqn:distance}, 
($\alpha_i$, $\delta_i$) and ($\alpha_0$, $\delta_0$) represent the RA and Dec of 
each CC and the centroid of the sample, respectively. By construction,
the X-axis is antiparallel to the RA axis, the 
Y-axis is parallel to the declination axis, and the Z-axis is towards 
the centroid of the SMC, i.e. increasing away from the observer. 

\subsection{Ages}

\begin{figure}
\includegraphics[width=8.5cm]{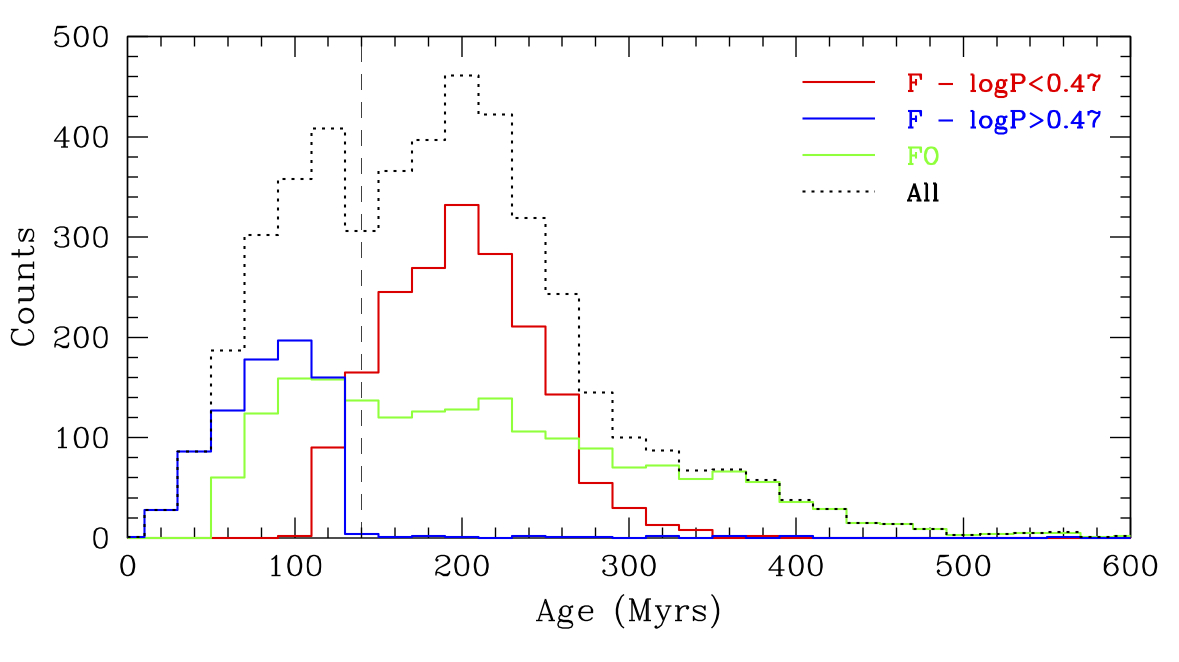}
\caption{Distribution of the individual ages for the CCs 
analysed in this work as obtained by means of the  $PAC$ relations 
\citep[after][]{Bono2005}. The vertical dashed line at 140 Myr is the
boundary between young and old CCs.
    \label{fig:histoAges}}
\end{figure}

An important ingredient for the following analysis is 
an estimate of the age of the CCs. This allow us to connect the pulsational properties of the CCs to
 those of the stellar population they belong to. This way we
can use the position and the 3D structure of the CCs to infer that of 
the young population of the SMC, with typical ages of 10-500 Myr.
The connection between the period and the age of a CC is rather
obvious, given that the $PL$ relations imply that at longer periods we
have brighter objects. In turn, the mass-luminosity relation
associated with  this range of masses means that brighter objects have higher masses
and hence younger ages. To estimate the ages of the investigated CCs we
adopted the Period--Age--Colour ($PAC$) relations for F and 1O pulsators at
Z=0.004 of \citet{Bono2005,Marconi2006} \citep[see also][for the possible effect of rotation on the
age of CCs]{Anderson2016}. The reliability of these relations has been confirmed in the
literature \citep[see e.g. the discussion in][]{Subramanian2015}. To 
calculate the  $(V-I)_0$ colour needed in the $PAC$ relations, we used
OGLE\,IV $V,I$ magnitudes and the reddening values listed in column 12 of
Table~\ref{table:averages}. The resulting ages are in the range
$\sim$10-900 Myr, with errors of the order of 5-30 Myr (larger errors
for increasing ages). 
The result of this procedure is shown in Fig.~\ref{fig:histoAges}.
As already noted by \citet{Subramanian2015,Jacy2016}, the distribution is almost
bimodal with two peaks at about 120($\pm$10) and 220($\pm10$) Myr. The bimodality of the
distribution is almost entirely caused by F pulsators which are clearly
separated depending on whether or not their  periods place them before or
after the break in the $PW$ relations, respectively. The 1O pulsators
 appear to have a smoother distribution, encompassing the whole
age range of the CC distribution.\footnote{This features can be explained
taking into account that  very young CCs can only be F pulsators, as
youth implies higher mass and luminosity, and hence higher periods. We
recall that the limiting periods for 1O pulsators is about 6 days.}  
Moreover, Fig.~\ref{fig:histoAges} shows two close main bursts of
CCs, and the most recent one seems to have formed less stars. 

We note that enhanced Star Formation Rate (SFR) around 200 Myr (to 0 Myr) was
shown in simulations \citep[see e.g. Fig. 7 in][]{Yoshizawa2003},
but they did not show two clear peaks. Also, there is no decline in SFR between
the two peaks. Hence, the observed bimodal age distribution among SMC
CCs represents an important constraint for the next generation of simulations.

It is useful for the following
analysis to divide the CCs into young and old samples, using 140 Myr
as limiting age. This value is somewhat different from that adopted by
\citet{Jacy2016}, as they used the \citet{Bono2005}'s Period-Age
relations instead of the more precise
$PAC$ ones as we did here. However, this difference is not
particularly significant for the following analysis.
 
\subsection{Photometric metallicities}

\begin{figure}
\includegraphics[width=8.5cm]{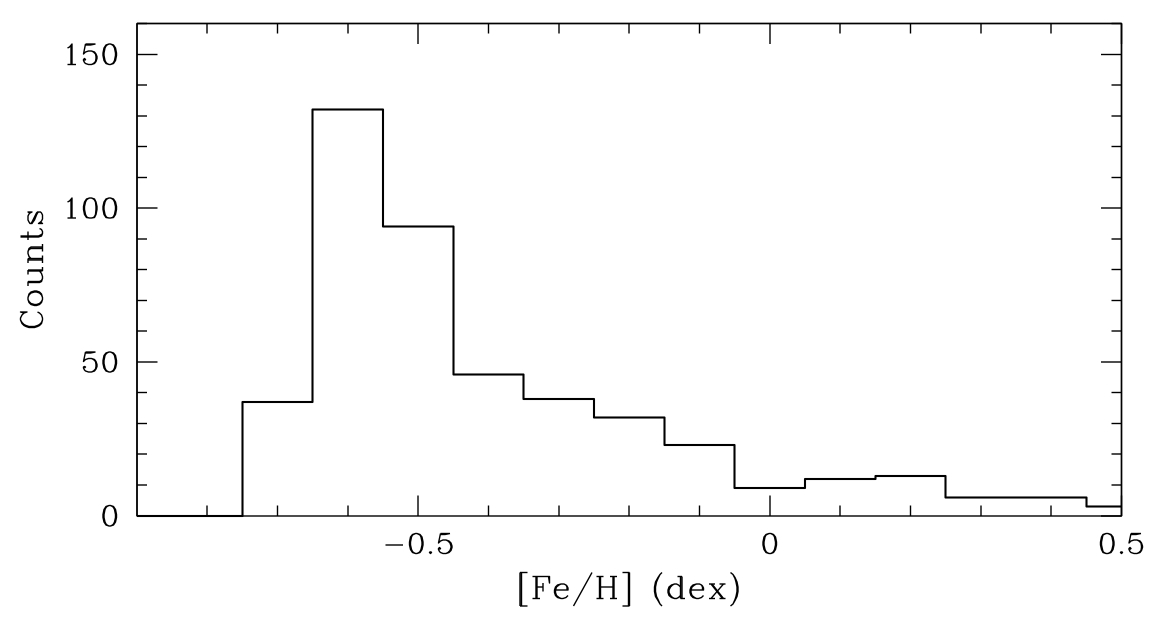}
\caption{Distribution of the individual photometric metallicities for the CCs 
analysed in this work. Note that the absolute [Fe/H] values can be
overestimated by $\sim$0.1 dex (see text). \label{fig:histoMet}}
\end{figure}

Metallicity, as well as age, is an important ingredient in the study of
 stellar populations. Extensive high-resolution spectroscopic surveys of CCs in the
SMC are not available as the current instruments do not allow to carry
out such measurements (this will be possible in the future with
e.g. 4MOST@VISTA). It is however possible to obtain 
an estimate of the CCs' iron content based on the shape (Fourier parameters)
of their light curves, as demonstrated by \citet{Klagyivik2013}, who
found multiband relationships connecting  [Fe/H]  with the $R_{21}$ or $R_{31}$
Fourier parameters. According to these authors, in the MCs their relations hold for F-mode
pulsators with periods in the range 2.5--4.0 days. 
To estimate the photometric [Fe/H] values we thus adopted the Fourier parameters $R_{31}$\footnote{$R_{31}$ is
  to be preferred to $R_{21}$ for lower metallicities, see Fig. 3 in \citet{Klagyivik2013}.} published by OGLE\,IV in the $I$
band and transformed it in the $V$ band, as according to
\citet{Klagyivik2013}, the $I$ band is less sensitive to metallicity.
The transformation used is the following: 

\begin{equation}
R_{31}(V)=(0.005\pm0.008)+(0.96\pm0.05)R_{31}(I) 
\label{eqn:r31}
\end{equation}

\noindent
with an rms=0.015 mag (Ripepi et al. in preparation). Finally, the
relations showed in Table 3 of \citet{Klagyivik2013} were 
used  to calculate the [Fe/H] values for 462 CCs in the SMC. 
A histogram of the results is shown in Fig.~\ref{fig:histoMet}. It
can be seen that there is a peak at about [Fe/H] =$-$0.6 dex, with a
tile of values reaching  [Fe/H] =$+$0.5 dex (11 stars with [Fe/H]
beyond this value are not shown). The minimum error on the individual
photometric [Fe/H] values is given by the rms of the $R_{31}$vs [Fe/H] relationship
adopted here, i.e. 0.082 mag. In addition, we have to consider the
random errors on the OGLE\,IV $R_{31}$ values, which are not available, and the uncertainty
 introduced by the use of Eq.~\ref{eqn:r31}. 
Moreover, the \citet{Klagyivik2013} relations have been calibrated
with Galactic CCs which are more metal rich than SMC ones, so that our
[Fe/H] values can be systematically overestimated by about 0.1 dex
\citet[see Fig. 3 of][]{Klagyivik2013}. Overall, a
conservative estimate of the total errors on the individual abundances is
thus of the order of 0.2 dex. We note however that the contribution of
random errors only should be smaller, so that relative individual abundances
among the SMC should be more robust than absolute values. As for the stars with
high values of [Fe/H] and in particular those with oversolar abundances,
we are not fully confident about the  reliability of these measures
that can be affected by wrong values of $R_{31}$. 

About 64\% and 73\% of the CCs have [Fe/H]$<-0.4$ and $<-0.3$
dex, respectively. The median of the distribution (to be preferred to
the average, given the highly asymmetric distribution) of the whole
sample is $-$0.50$\pm$0.16 dex. This value become $-$0.56$\pm$0.06 dex $-$0.52$\pm$0.08 dex if we
consider only pulsators with [Fe/H]$<-0.4$ and $<-0.3$ dex,
respectively.  

These values are somewhat higher than the
estimate by \citet{Romaniello2008}    [Fe/H] =$-$0.75 dex ($\sigma
\sim$ 0.08 dex) on the basis of a sample of 14 CCs observed 
with high-resolution spectroscopy, or that by \citet{Piatti2013}, [Fe/H] =$-$0.70 dex ($\sigma
\sim$ 0.15 dex) based on Washington photometry of field stars (present
day metallicity, their Fig. 6). The discrepancies are of the order of
0.2 and 0.15 dex for the whole and the more metal poor samples,
respectively. Hence, formally, our values agree with literatures'
within $\sim 1 \sigma$.
However, as already remarked above, we expected to overestimate the
abundances by some 0.1 dex due to the lack of metal poor calibrators in
\citet{Klagyivik2013}'s work, therefore the slight discrepancy
observed is real.  

\begin{figure}
\includegraphics[width=8.5cm]{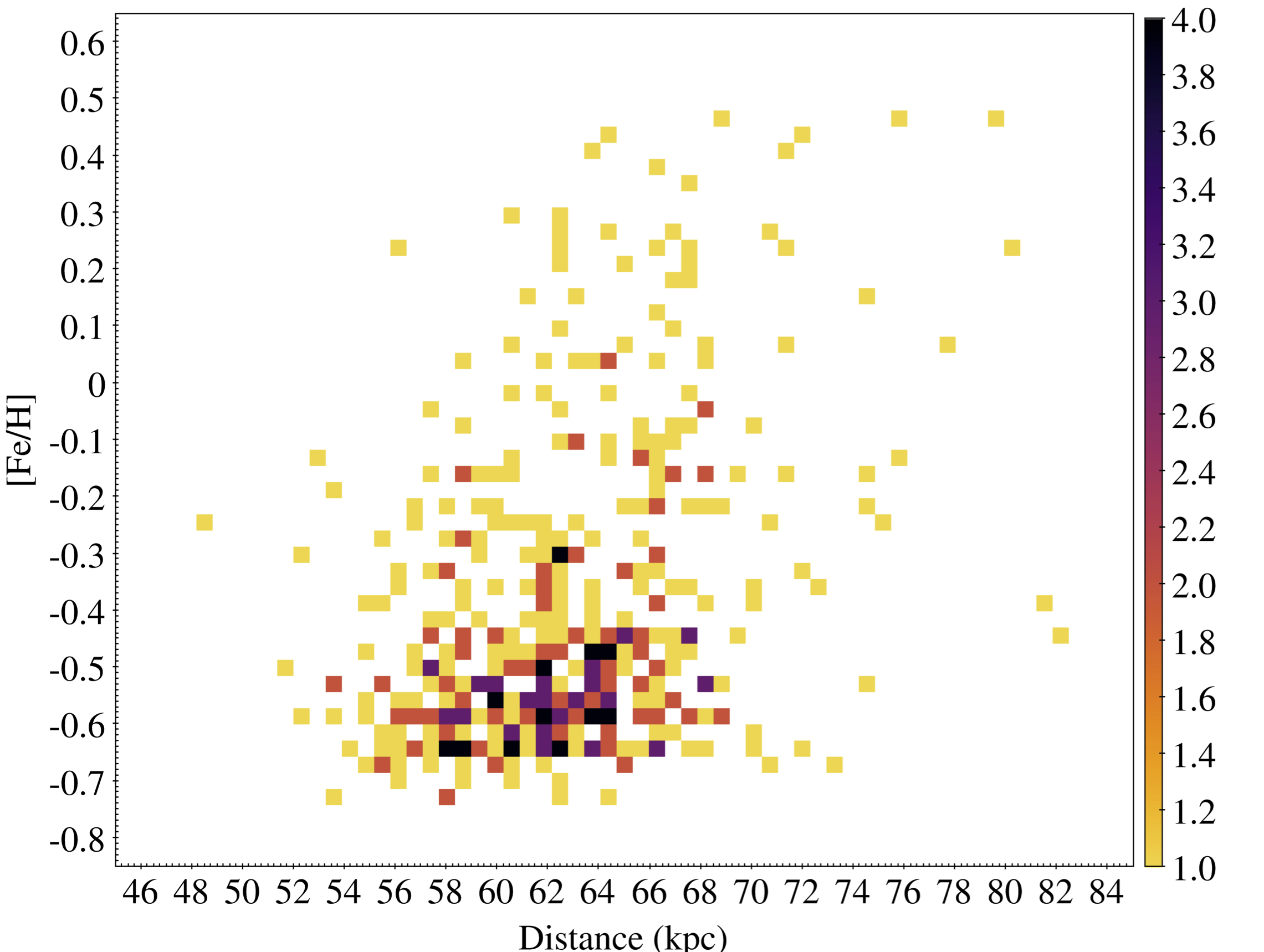}
\caption{Photometric metallicities as a function of distance from the
  Sun. The centre of the SMC is placed at 63 kpc. The colour bar shows
  the number of objects in each bi-dimensional bin. Note that the high
  value of [Fe/H] measured for a minority of CCs can be affected by 
a large uncertainty (see text). \label{fig:metDist}}
\end{figure}

In any case, it is reasonable to assume that 
the relative abundance differences are usable. Therefore, we searched
for  possible trends between metallicity and position in the
SMC. Figure~\ref{fig:metDist} displays the photometric  [Fe/H]  values as
a function of the distance from us. 
The figure shows that CCs closer than roughly 60 kpc seem preferentially more metal poor
than objects between 60 and and 70 kpc, i.e. roughly around the centre
of SMC at 63 kpc. This could imply that the stripped material from which the more
external CCs formed was relatively metal poor, while the one closest
to the center was enriched between the time when there was a tidal
interaction that brought out the material and the time when the cepheids
were formed.
However we remind our warning about the possible
large uncertainties affecting high abundance evaluations.  Indeed, 
considering the sample of objects with [Fe/H]$<-0.4$ dex, the
distribution appears rather uniform with distance.

\begin{figure*}
\includegraphics[width=12cm]{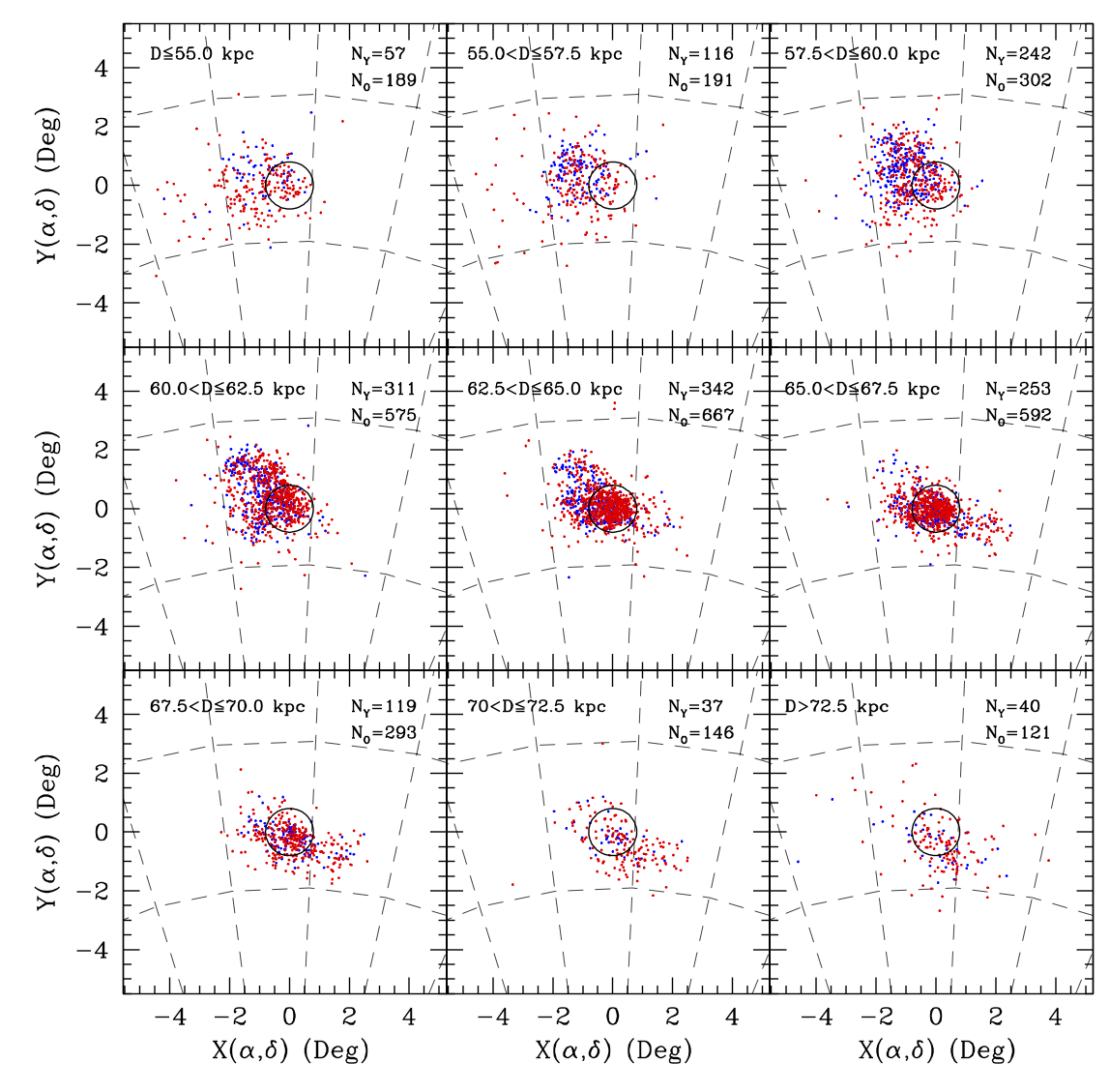}
\caption{Distribution of SMC CCs on the sky at varying distances. Blue  
  and red dots show pulsators with age$<$140 Myr and age$\ge$140  
  Myr, respectively. The number of pulsators within or beyond this
  treshold are labelled in each panel with N$_{\rm Y}$ and N$_{\rm
    O}$, respectively. A circle centred on (0,0)  
  coordinates and with radius=0.8 kpc has been drawn in each panel for
  reference. Note that in this zenithal  
  equidistant projection, N is up and E  
  is to the left (the X axis is antiparallel to the $\alpha$ axis). The centre is at  
  $\alpha_0$=12.54 deg; $\delta_0=-73.11$ deg. \label{fig:2col}}
\end{figure*}

\subsection{Distribution of CCs on the sky}

We divided our sample of CCs according to
their distances, adopting an interval of 2.5 kpc between adjacent
subsamples. This value was chosen because it is larger than the
individual errors on the distances. It is also sufficiently small to allow us to investigate in detail the
distribution of CCs in the sky including a statistically significant
number of stars in each interval. The result of this exercise is shown
in Fig.~\ref{fig:2col} where we show the distribution of the CCs in
the sky \citep[following][we adopted a cartesian zenithal equidistant projection]{vandermarel2001}
at varying distances and different colours for ages
$<$ 140 Myr (blue) and  $\ge$ 140  Myr (red). The relative distances
shown in the figure have been obtained with the $PW(V,K_\mathrm{s})$,
but we verified that the use of the $PW(J,K_\mathrm{s})$ leads to the
same results.

An analysis of the figure confirms that the SMC is highly
elongated along the LOS. However, the exquisite accuracy of
our relative distance determinations allows us to detect more details
about the structure of the galaxy. First, using the central circle
with radius=0.8 kpc as reference, it can be easily seen that
the galaxy is elongated roughly from east (E) towards the 
west/south--west (W/SW) direction, with the stars in the latter position the farthest from
us. The structure of the SMC appears approximately round closer to us (for distances
smaller than $\sim$60 kpc) and becomes progressively
asymmetric going to greater distances, assuming an asymmetric distribution with
 inclination in the north--west (NW)--south--east (SE) direction, that can be clearly seen beyond
$\sim$ 65 kpc. Around the distance of the centre of the galaxy,
i.e. 63 kpc, there is a bulk of CCs with approximately
round shape (and radius of about 0.8 kpc, as shown by the circle in
the figure) that is visible until 67.5 kpc. A strong structure
is present in the NE direction between 57.5 and 65 kpc 
\citep[similar to the northern structure discussed by][]{Jacy2016}. 
At 65 kpc this NE structure disappears and 
a concentration of stars in the SW direction becomes visible. This feature is present 
until and beyond 72.5 kpc. There are apparently no clear
trends with the age of the pulsators, apart from a striking feature present
for distances smaller than 55 kpc (uppermost-left panel in Fig.~\ref{fig:2col})
in the easternmost part of the whole galaxy distribution. The
pulsators in this isolated region are predominantly older than 140
Myr. We will speculate further on the age trends in the next section. 

\begin{figure*}
\includegraphics[width=16cm]{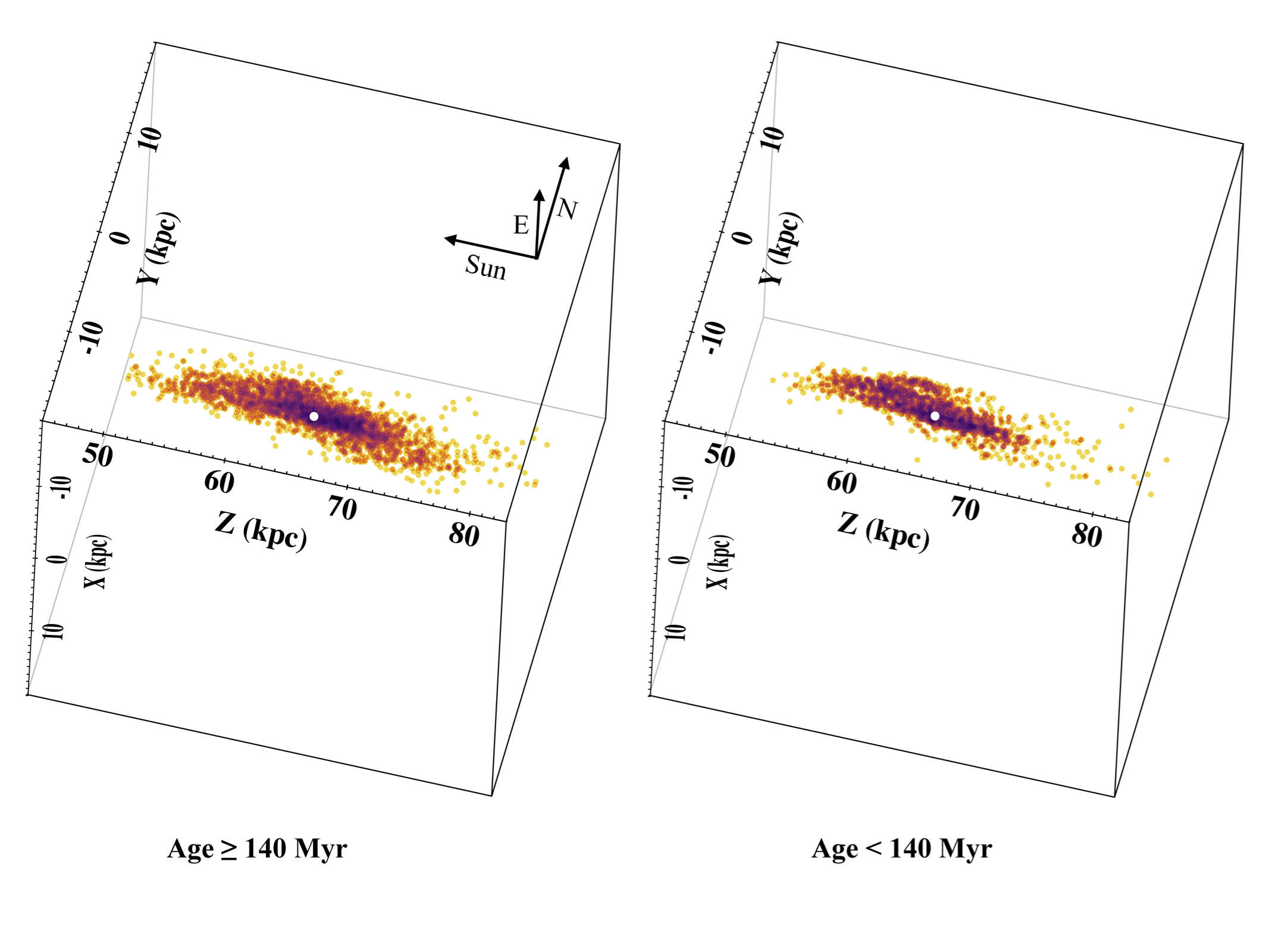}
\caption{Three--Dimensional Cartesian distribution of the SMC CCs for the 
  labelled ages. The colour scale is proportional to the star density. In both panels a white filled circle shows the 
  approximate centroid of the whole sample of CCs adopted in this work 
  ($\alpha_0$=12.54 deg; $\delta_0=-73.11$ deg). \label{fig:plot3D}}
\end{figure*}

\begin{figure*}
\includegraphics[width=12cm]{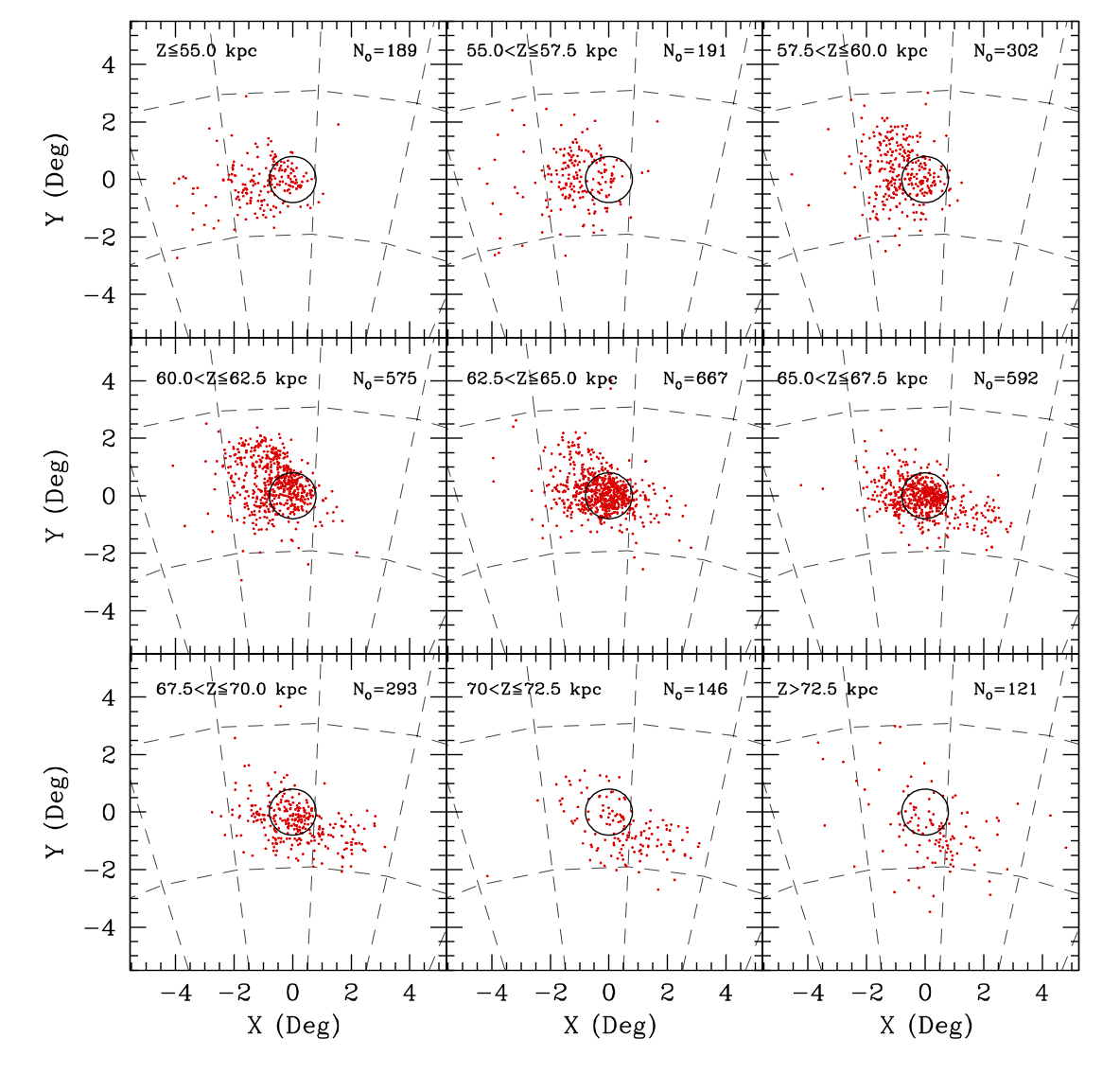}
\caption{Cartesian X,Y plane at varying Z for the old sample (age
  $\ge$ 140 Myr). As in Fig.~\ref{fig:2col}, each panel shows a circle centred on (0,0)  
  coordinates and with radius=0.8 kpc. \label{fig:xy_old}}
\end{figure*}

\begin{figure*}
\includegraphics[width=12cm]{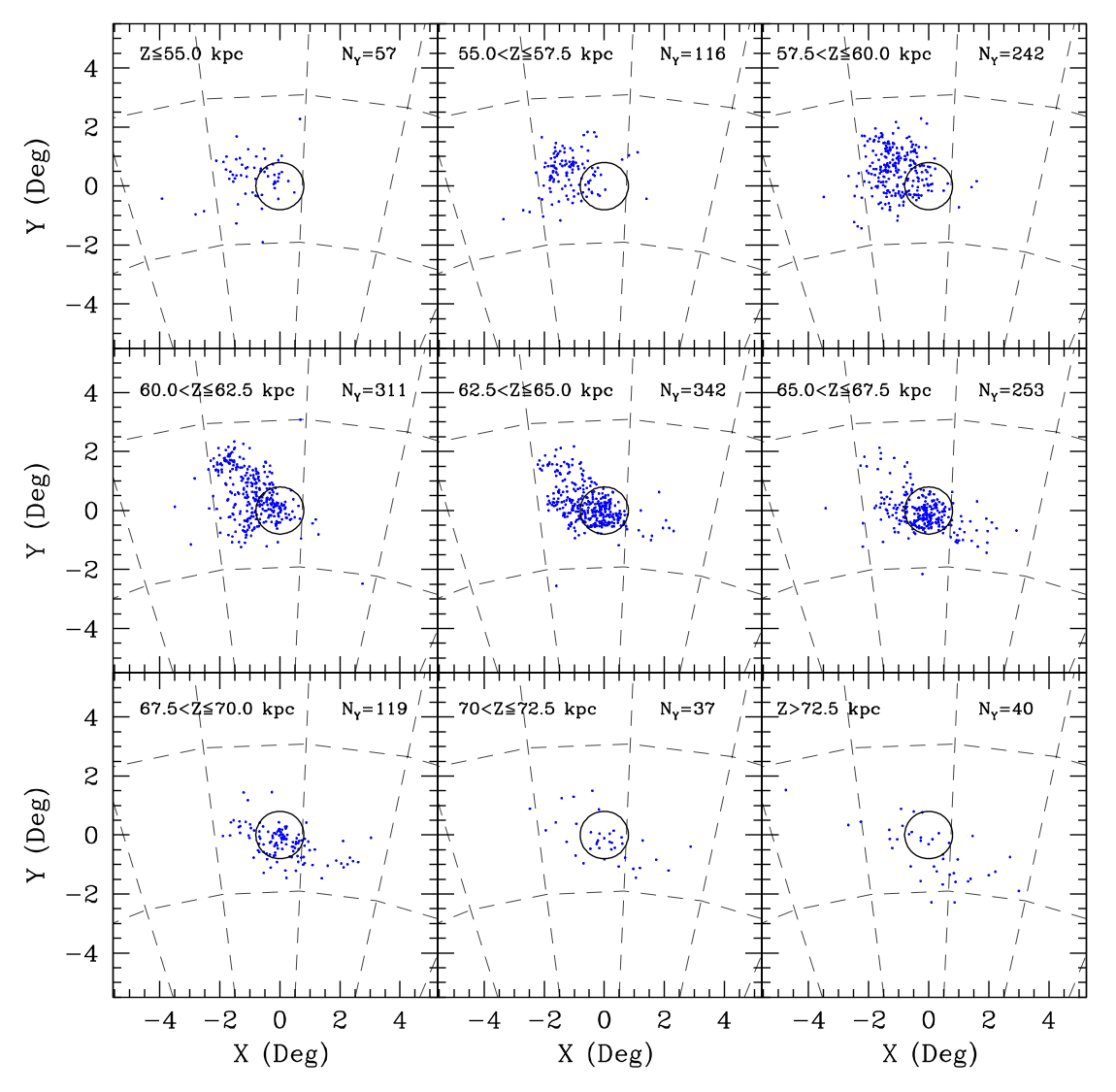}
\caption{As in Fig.~\ref{fig:xy_old} for the young sample (age
  $<$140 Myr). \label{fig:xy_young}}
\end{figure*}

\begin{figure*}
\includegraphics[width=12cm]{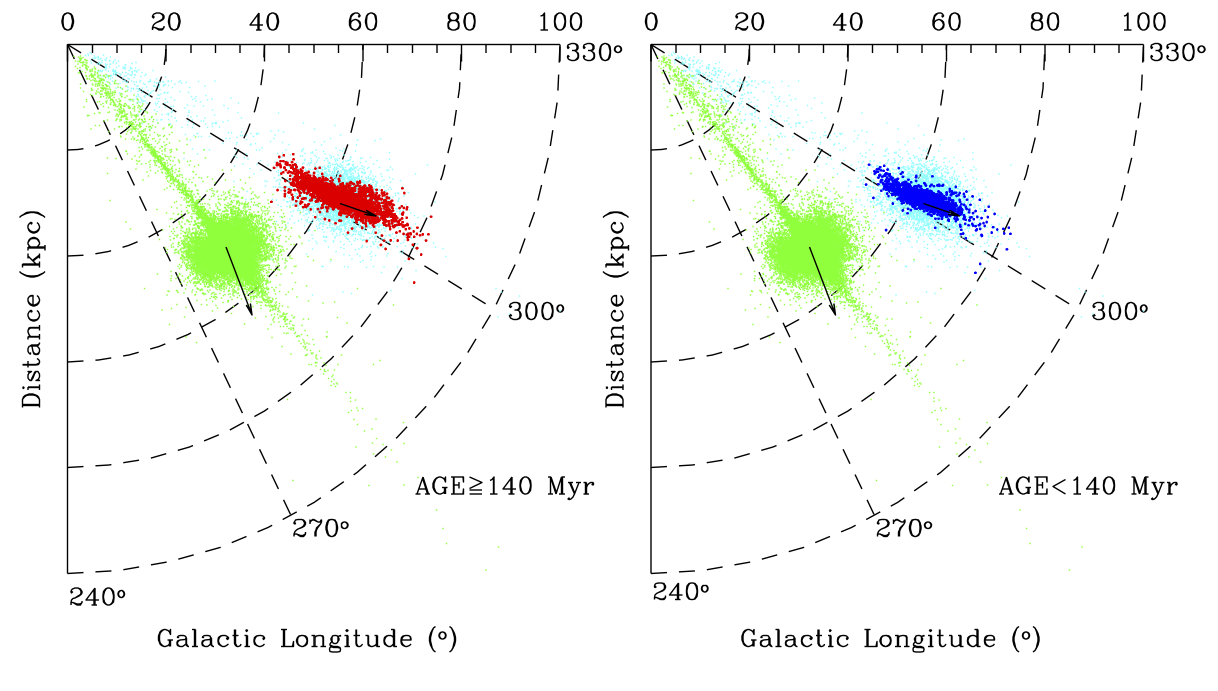}
\caption{Polar projection of Galactic longitude coordinates for the 
  old (red dots) and young (blue dots) CC samples in the SMC. For 
  comparison, the ab--type RR Lyrae stars in the SMC (cyan dots) and in 
  the LMC (green dots) are shown. The black arrows show the directions 
  of motion of the two galaxies (the position they will reach 
  in 50 Myr), according to the proper motion measurements of
  \citet{vandermarel2016}. We note that according to \citet{Jacy2017}, the straight lines of stars
  connecting between R=0 kpc to R=80 kpc in both the LMC and SMC are 
either true Galactic RR Lyrae variables along the LOS or (in large
majority) pulsators whose distances are strongly affected by blending/crowding or other photometric errors.
\label{fig:polar}}
\end{figure*}

\subsection{Distribution of CCs in 3D cartesian space}
\label{questaSection}

The distribution of SMC CCs in cartesian 3D space is shown in
Fig.~\ref{fig:plot3D}, where the left and right panels show the sample of
CCs with age $\ge$140 Myr and age $<$140 Myr, respectively. The scales
along the three axes are the same, so that the elongated nature of the 
distribution of CCs in the SMC can be easily seen. As already noted
by \citet{Jacy2016} and \citet{Scowcroft2016}, the distribution of CCs in the
SMC is not planar, but rather ellipsoidal, irregular and somewhat
fragmented. 
Some of them are more evident in the old sample (age
$\ge$140 Myr), whereas others stand out clearly in the young sample
(age $<$140 Myr). These features can be discussed more in detail by 
analysing the projection in the cartesian X,Y plane for different
values of Z, as shown in Fig.~\ref{fig:xy_old} and ~\ref{fig:xy_young} for
the old and young samples, respectively. An analysis of these figures
reveals a close similarity with Fig.~\ref{fig:2col},
indicating that the elongation of the SMC is almost along the LOS. 
The presence of an off--centre eastern structure for Z of less than 
approximately 60 kpc is confirmed. This structure is almost completely 
populated by CCs of the old sample, which also shows an extreme
E/SE component, almost absent from the young sample. 
Beyond 60 kpc the general distribution of CCs is more centred, and the most evident
substructure is shifted towards the NW. Between 60.0 and 62.5 kpc,
this north-eastern substructure is more evident in the young sample,
which 
seems to show a possible trimodal distribution in this interval of
distance. Looking at Fig.~\ref{fig:plot3D} right panel and
Fig.~\ref{fig:xy_young} mid-leftmost panel, the possible centres of
this trimodal distribution are located approximately around ($-$0.5,0.0), ($-$1.0,$-$1.0) and ($-$1.5,$-$2) kpc,
respectively. Some signs of this feature are present until Z$\sim$65.0
kpc in the young sample, whereas the old one is almost clustered in
the central 1 kpc. Beyond Z=65.0 kpc the young sample is almost
centred without significant substructures. On the contrary, the
old sample shows a remarkable SW structure extending from
Z$\sim$65.0 kpc to Z$\sim$72.5 kpc and beyond \citep[see
also][]{Jacy2016}. Summarizing, the old sample appears to be more
elongated than the young one both in the E  (closer to us) and
in the SW (farther from us) directions \citep[this 
is in contrast with][who found that the stars closer to us are
also the youngest]{Jacy2016}. The young sample is
thus more compact than the old one, as there are very few young CCs closer than 55 kpc and
beyond Z$\sim$70 kpc. Its location is apparently on
average more off--centre (see Fig.~\ref{fig:xy_old} and
Fig.~\ref{fig:xy_young} for Z between  60 and 67.5 kpc)
with respect to the bulk of old CCs. Taking into account that 
 the LMC is located in the E/NE (depending on the distance)
 direction with respect to the SMC, we can use this empirical
 evidence to suggest a tentative explanation for this complex
 distribution. Thus, the population to which the old CCs belong underwent a strong interaction with
 the LMC when they had already been formed (i.e. $>$ 140 Myr ago) and the
 signatures of this event are: i) the distribution of stars closer than
 $\sim$60 kpc heavily off-centred in the E/NE direction,
 i.e. towards the LMC; ii) the spread of old pulsators in the opposite
 direction (SW), as expected in case of strong tidal
 interactions (see also Sect. ~\ref{discussion}). On the contrary the young CCs were probably formed
 after that event, but part of the gas/material from which they formed
 was already subject to this gravitational interaction. 
This would explain the strong off--centred (eastern) position of the
 majority of the young CCs until Z$\sim$60 kpc, and the NE
 substructure(s) clearly visible among young pulsators for Z 
 between $\sim$60 and $\sim$65 kpc. In this scenario, the lack of young pulsators for
 Z$>$70 kpc can be explained by hypothesising that the gas/material
 tidally stripped in this direction, if any, did not possess sufficient density
 to trigger strong star formation, because originally more concentrated
 in the SMC's body \citep[in this region
 there is no strong recent star formation, see e.g.][]{Rubele2015}. 

\begin{figure*}
\includegraphics[width=12cm]{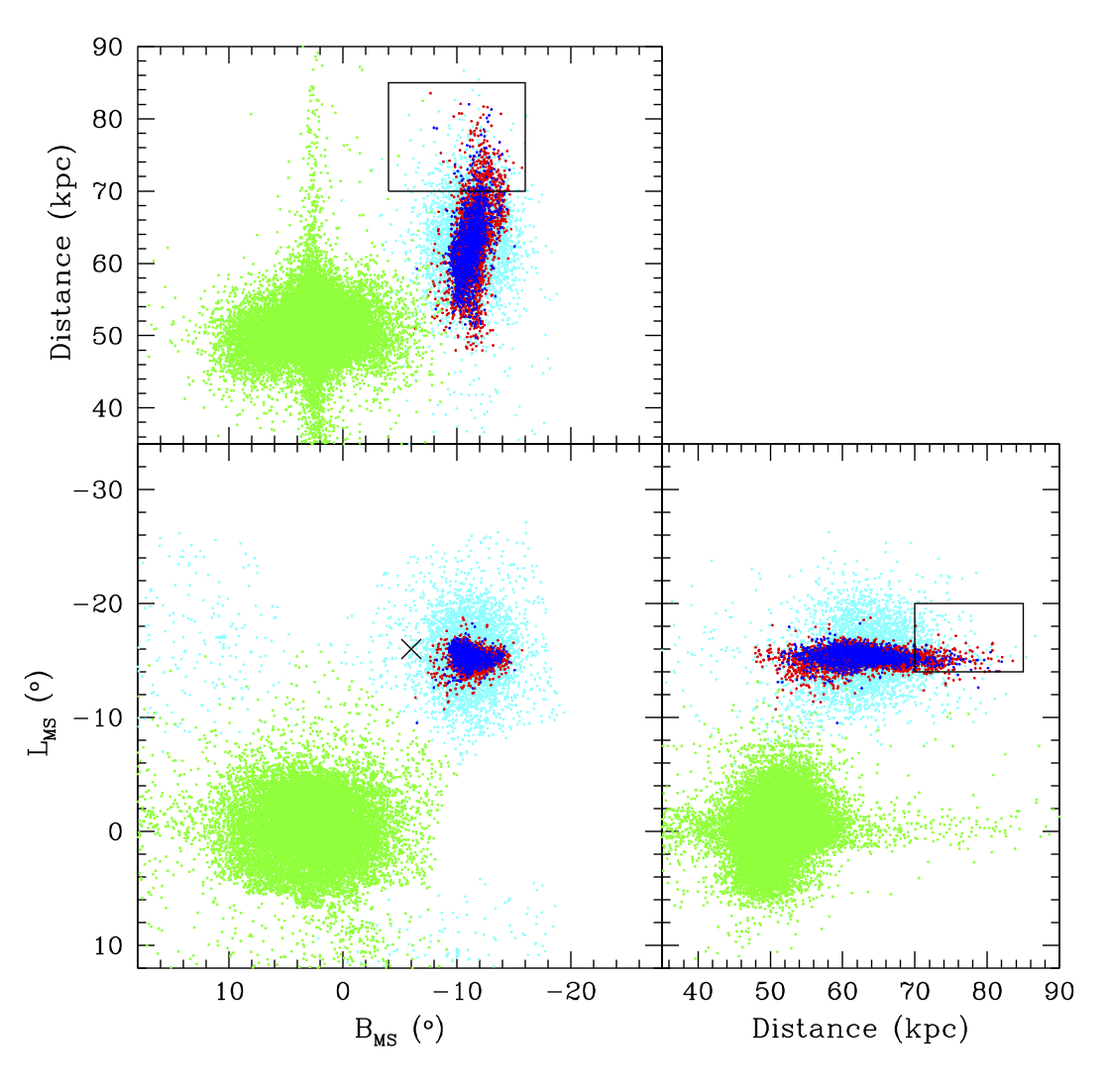}
\caption{Position of LMC and SMC  RR Lyrae as well as of young and old
  CCs in the SMC (same colour coding as in Fig~\ref{fig:polar}) in Magellanic coordinates (bottom--left) or
  Magellanic longitude/latitude vs distance (bottom--right and
  top--left, respectively). This figure is similar to Fig. 8 of 
  \citet{Diaz2012}, whose prediction can be directly compared. The
  approximate location of the predicted Counter--Bridge are shown with
  a diagonal cross (centre) in the bottom--left panel and with boxes
  in the top--left and bottom-right
  panels. The colour code of the data points is the same as in Fig.~\ref{fig:polar}.
\label{fig:magellanicCoord}}
\end{figure*}

 This scenario can be further appreciated by plotting the CC distribution 
in terms of Galactic longitude vs distance, see 
Fig.~\ref{fig:polar}. In addition, to show a more complete picture of the
Magellanic system, we plotted the positions of the ab-type RR Lyrae stars in
the same plane for both the LMC (green dots) and SMC (cyan dots). The
distances of the RR Lyrae variables in both galaxies were derived 
from the optical OGLE\,IV photometry as explained in Appendix~\ref{appendixB}
The figure also shows with black arrows the direction
  of motion of the two galaxies (i.e. the position they will reach
  in 50 Myr), according to the proper motion 
measurements of \citet{vandermarel2016}. This figure suggests that: 
i) in the SMC, the distributions of the
CCs and RR Lyrae have the same elongation, but different shapes, in the
sense that the RR Lyrae stars define a larger spheroidal
or ellipsoidal volume \citep[see][and references
therein]{Jacy2017,Muraveva2017}; 
ii) it is confirmed that the distribution of the young CCs is more compact
than the old CCs one; iii)  the elongation of the SMC is approximately
oriented in the direction of the MW,
but there are clear hints of a distortion, visible especially
for old pulsators in the direction of the LMC in the distance range
from 50 to 60 kpc; 
iv) the elongation of the SMC CCs is consistent with the direction of
the SMC motion. 

\section{Discussion}
\label{discussion}

Previous results on the three--dimensional structure of SMC CCs,
obtained by \citet{Haschke2012} and \citet{Subramanian2015}, based on 
OGLE\,III data, were based on the assumption that the young stars are
more likely to have a disk distribution. This assumption was based on
the spectroscopic study of young stars by \citet{Evans2008} and
\citet{Dobbie2014}. The latter authors suggested that 
the red giant distribution in the SMC is better explained by a rotating disk of 
inclination  and position
angle of the line of nodes similar to that of the parameters of the
disk found using CCs. These results were contradicted by the recent 
work of \citet{vandermarel2016}, who used HST and $Gaia$ proper
motions in the SMC, finding no significant rotation (i.e. the signature of a distribution on a disk) for young 
stars, as that seen in previous studies \citep[but][state that the next data release of 
$Gaia$ is needed to obtain conclusive results]{vandermarel2016}. 
In any case, in the disk model by \citet{Haschke2012} and 
\citet{Subramanian2015} the large elongation along LOS is explained
 as due to a large inclination of the disk. 

However, these outcomes were questioned by recent
works based on more accurate measurements 
\citep{Scowcroft2016}, or more extended samples \citep{Jacy2016} of CCs.
Our results, as presented in the previous Section are in substantial
agreement with these two latter works. Indeed, we support their conclusions
concerning: i) the SMC CC distribution is not planar, but elongated
approximating an ellipsoidal shape with
significant substructures; ii) a different distribution of the young and
old CC samples \citep{Jacy2016}  and the rotation of the distribution in the direction
of the LMC \citep{Scowcroft2016}. However, the high precision of our measurements
exploiting the NIR bands allowed us to carry out the finest
characterization of the three--dimensional distribution of SMC CCs.  

Very recently \citet{Subramanian2017}, using VMC survey data for intermediate-age
RC stars, found a foreground population at about
12$\pm$2 kpc (with respect to the SMC centre) from the inner ($r \sim 2 \degr$) to the outer ($r \sim 4 \degr$)
regions in the eastern SMC. They propose that these stars have been
tidally stripped from the SMC during the most recent LMC--SMC
encounter and thus offer evidence of the tidal origin of the MB.  An inspection of
Fig.~\ref{fig:plot3D}--\ref{fig:polar} reveals that our results do not
support such a bimodal distribution of CCs in the eastern
regions of the SMC, even for the old CC population, which is expected to
be affected by the close encounter with the LMC (the younger one
should have formed CCs after that episode) as for the
intermediate--age RC stars. However, we have only a few CCs at radii larger than $2 \degr$
in the eastern direction (see the top--leftmost panel of
Fig.~\ref{fig:2col}), where, according to \citet{Subramanian2017}, the
effect of the bimodality of RC stars is maximum. Moreover, these
easternmost CCs are at closer distances to us than the SMC centroid
($>$8 kpc closer).  
On the basis of these considerations, we can conclude that the results from CCs and RC are
consistent.

The results presented in the previous section can also be compared with the
predictions of \citet{Diaz2012}. To this end we constructed a figure similar
to their Fig. 8, but using our data for RR Lyrae stars in the LMC and
SMC, as well as for young and old CCs in the SMC. The result of this
exercise is shown in Fig.~\ref{fig:magellanicCoord}, where, following
\citet{Diaz2012}, we adopted the Magellanic Coordinate System ($L_{\rm MS}$,$B_{\rm MS}$)
introduced by \citet{Nidever2008}. It is worth
recalling that one of the most prominent features in \citet{Diaz2012}'s
predictions, is the existence of the Counter--Bridge (CB) in opposite
direction with respect to the Magellanic Bridge (MB).
The presence of CB in the western region was also discussed briefly in
\citet{Subramanian2015}. There the CB counter-parts are
identified as those which are out of the plane and behind the
disk, i.e. CCs at distances of 70-80 kpc.
In Fig.~\ref{fig:magellanicCoord} the CB is expected to be located at
($L_{\rm MS}$,$B_{\rm MS}$)$\approx$($-16\degr,-6\degr$) in the bottom--left
panel of Fig.~\ref{fig:magellanicCoord}, between  $-4 \degr \la B_{\rm MS}
\la -16 \degr$ in the top--left panel and $-14 \degr \la L_{\rm MS}
\la -20 \degr$ in the bottom--right  panel. These locations have been
approximately highlighted in the figure to make the comparison 
with \citet{Diaz2012}'s Fig. 8 easier. An inspection of
Fig.~\ref{fig:magellanicCoord}, thus reveals that the regions where the CB is
expected are partially populated (both in $B_{\rm MS}$ vs distance and
$L_{\rm MS}$ vs distance diagrams) by old CCs as well as RR Lyrae, whereas young
CCs are more centrally concentrated. This finding is in agreement with
our scenario outlined in Sect.~\ref{questaSection} and seems to
confirm \citet{Diaz2012}'s model predictions.

Figure~\ref{fig:magellanicCoord} shows additional interesting
features showing once again the effects of the interaction between the
MCs: i) the already known fact that there is no actual separation between the
two MCs is clearly visible in these plots; ii) the distribution of RR
Lyrae and CCs is clearly disturbed by the presence of the LMC; iii) the
distribution of young CCs is different from that of the old ones,
being more concentrated and showing a larger rotation towards the LMC,
especially in the $B_{\rm MS}$ vs distance plane.

These features are also in fair agreement with Model 2 of \citet{Besla2012},
based on the hypothesis of a direct collision between the MCs, leaving
a trail of stars between the MCs. This presence is evidenced by the
continuous distribution of pulsators in Fig.~\ref{fig:magellanicCoord} (as well as in
Fig.~\ref{fig:polar}). \citet{Besla2012}'s model also
predicts star formation along the MB at an epoch compatible with the
age of the CCs (e.g. 100-300 Myr). From our analysis, the CCs closer
to the MB appear to be the oldest ones but this does not appear in
contrast with this models, given the wide range of
ages encompassed by \citet{Besla2012}'s prediction.

\section{Conclusions}

In this paper we have presented the VMC survey's light curves for 717
CCs in the SMC. These data complete and complement our previous 
Paper I, resulting in a sample of 4793 CCs in SMC for which we provide   
here $Y$, $J$, and $K_\mathrm{s}$ average magnitudes, amplitudes and
relative errors, calculated as in Paper I. Our work represents the
complement to the optical study of CCs in the SMC carried out by the
OGLE group \citep{Soszynski2015b,Jacy2016}.  

The intensity-averaged magnitudes in the VISTA $Y$, $J$, and
$K_\mathrm{s}$ filters have been complemented with optical $V$-band
data and periods to construct multi-filter $PL$ and $PW$
relations for SMC CCs. In particular, the $PL$ and 
$PW$ relations in the $V, J$, and $K_\mathrm{s}$ bands for  
F- and 1O-mode SMC CCs presented in this work are the most accurate to
date, based on well- or moderately well-sampled light
curves in $K_\mathrm{s}$ and $J$, respectively.

 We used these 
 relations to estimate accurate individual CC distances in the SMC
 covering an area of more than 40 deg$^2$. Adopting literature relations,
 we estimated ages and metallicities for the majority of the
 investigated pulsators, finding that: i) the age distribution is bimodal,
 with two peaks at 120 and 220 Myr and a break at 140 Myr; ii) the
more metal-rich CCs appear to be located closer to the centre of the
galaxy.  

We provided the most accurate 3D distribution of the CCs in the SMC to
date. We conclude that,
in agreement with the literature, the distribution of the CCs is not planar but
significantly elongated over more than 25-30 kpc approximately in the
E/NE towards SW direction. Our exquisite photometry allowed us to describe in
great detail the substructures present in the SMC. In more detail, the young and old CC 
populations in the SMC seem to show  different distributions. The
distribution of the old CCs is more elongated with respect to the
young one that is more centrally concentrated. Both old and young
stars show strong off--centre structures due to the history of
interaction with the LMC.  

A possible scenario able to explain the present time 3D geometry of
the CCs in the SMC is the following: the population, to which the old CCs
belong, underwent a strong interaction with the 
 LMC when they were already formed (i.e. $>$ 140 Myr ago). The
 signatures of this event are: i) the distribution of stars closer than
 $\sim$60 kpc heavily off-centred in the E/NE direction,
 i.e. towards the LMC; ii) the spread of old pulsators in the opposite
 direction (SE), as expected for strong tidal
 interactions. On the contrary, the young CCs were probably formed
 after that event, but part of the gas/material from which they formed
 was already subject to gravitational interaction. 
This would explain the strong off--centre (eastern) position of the
 majority of the young CCs until Z$\sim$60 kpc, and the NE
 substructure(s) clearly visible among young pulsators for Z 
 between $\sim$60 and $\sim$65 kpc. In this scenario, the lack of
 young pulsators at 
 Z$>$70 kpc can be explained by hypothesising that the gas/material
 tidally stripped in this direction did not have sufficient density
 to trigger strong star formation \citep[in this region
 there is no strong recent star formation, see e.g.][]{Rubele2015}. 

In addition, our results indicate that the shape of the SMC is elongated and
somewhat rotated towards the  LMC in the region closer to this galaxy,
as expected from the hypothesis of a direct collision or a close
encounter some 200 Myr ago. 

A detailed comparison between our observations and the \citet{Diaz2012}
predictions showed a fairly good overall agreement, with the first
confirmation of  the presence of the CB, remaining after the last close
LMC--SMC encounter. 
It appears now clear that a new generation of models of the
MS is needed to explain the detailed and complex
structure of the young SMC component traced by its CC population, and
described in detail in this paper. 

Finally, we do not think that much progress in the knowledge of the
geometry of the SMC will be possible in the future from the photometry of CC
variables only. Though, information is lacking 
about proper motions, radial velocities and abundances for these
stars. Proper motions will be the subject of future works in the context
of the VMC survey itself \citep[see e.g.][]{Cioni2016} or
available for not too crowded objects by means of the {\it Gaia} satellite
measurements. Massive spectroscopic surveys 
of the pulsating variables will be possible within a few years thanks
to instruments such as 4MOST@VISTA and MOONS@VLT. 
The sum of this complementary information will eventually allow us to
unveil with great confidence the 
``true''  story of the formation and evolution of the SMC and MS as a
whole.

\section*{Acknowledgements}

We thank D.L. Nidever for kindly providing us with the code to
transform Galactic to Magellanic coordinates.

This paper is based on observations obtained with the ESO/VISTA telescope
located at Paranal (Chile).  We thank the UK's VISTA Data Flow System
comprising the VISTA pipeline at CASU and the VISTA Science Archive at
Wide Field Astronomy Unit (Edinburgh; WFAU) for providing calibrated
data products supported by the STFC.

This project has received funding from the European Research Council
(ERC) under the European Union's Horizon 2020 research and innovation
programme (grant agreement No. 682115). M.-R. Cioni acknowledges
support from the UK's Science and Technoloy Facility Council [grant
number ST/M001008/1].

R. d. G. is grateful for research support from the National Natural
Science Foundation of China through grants U1631102, 11373010 and
11633005.

S. Subramanian acknowledges research funding support from the Chinese
Postdoctoral Science Foundation (grant number 2016M590013).








\appendix

\section{Light Curves}
\label{appendixA}

\begin{figure*}
\includegraphics[width=16cm]{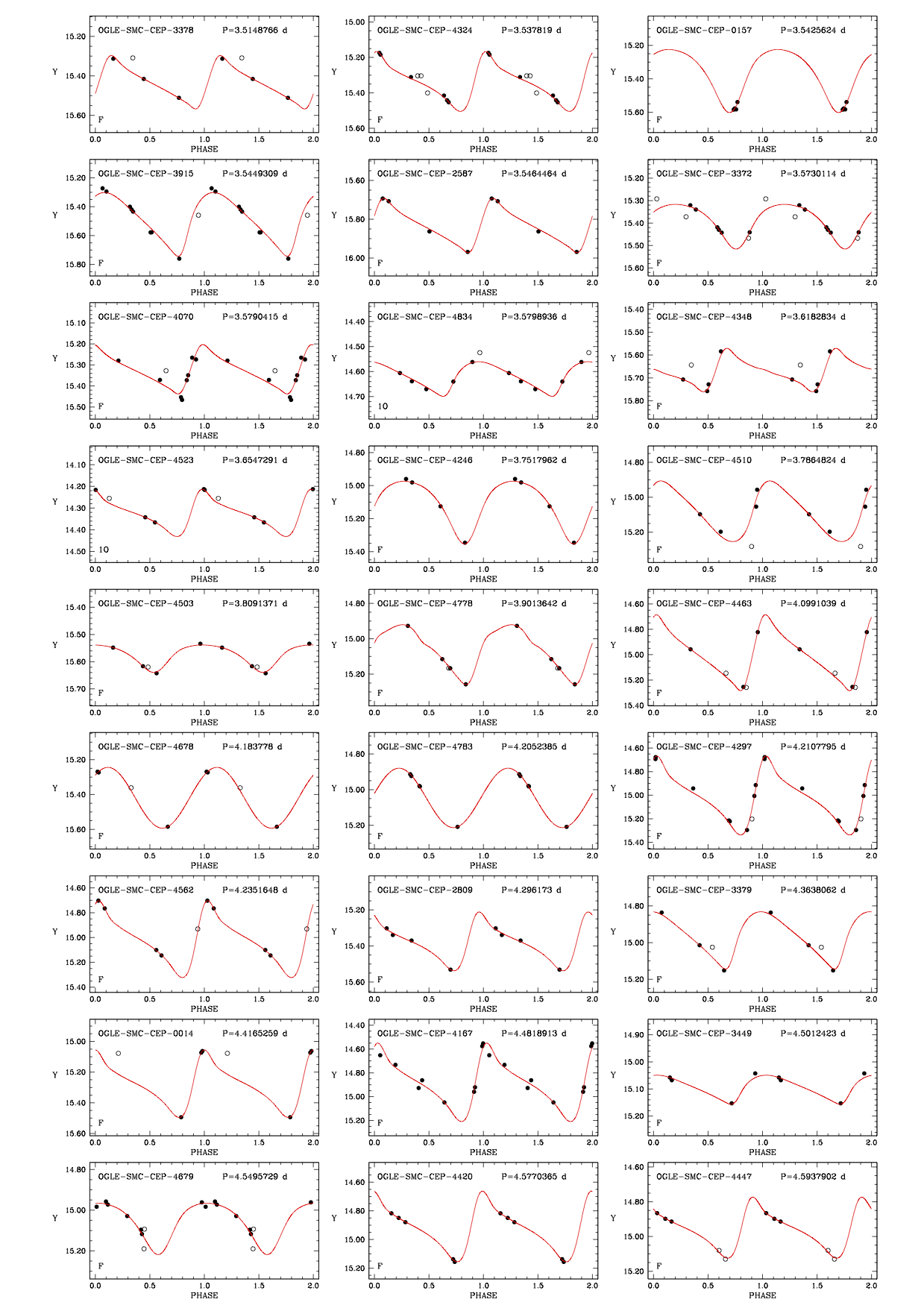}
\caption{$Y$-band light curves for the 757 CCs with new data with 
  respect to Paper I (see text). Stars are displayed in order of increasing 
period. Filled and open circles represent phase points used or not 
used in the fitting procedure, respectively (see Paper I). Solid lines represent 
best-fitting templates to the data (see text). In each panel we 
report OGLE's identification number, mode of pulsation and period.} 
\label{fig:CLY}
\end{figure*}

\begin{figure*}
\includegraphics[width=16cm]{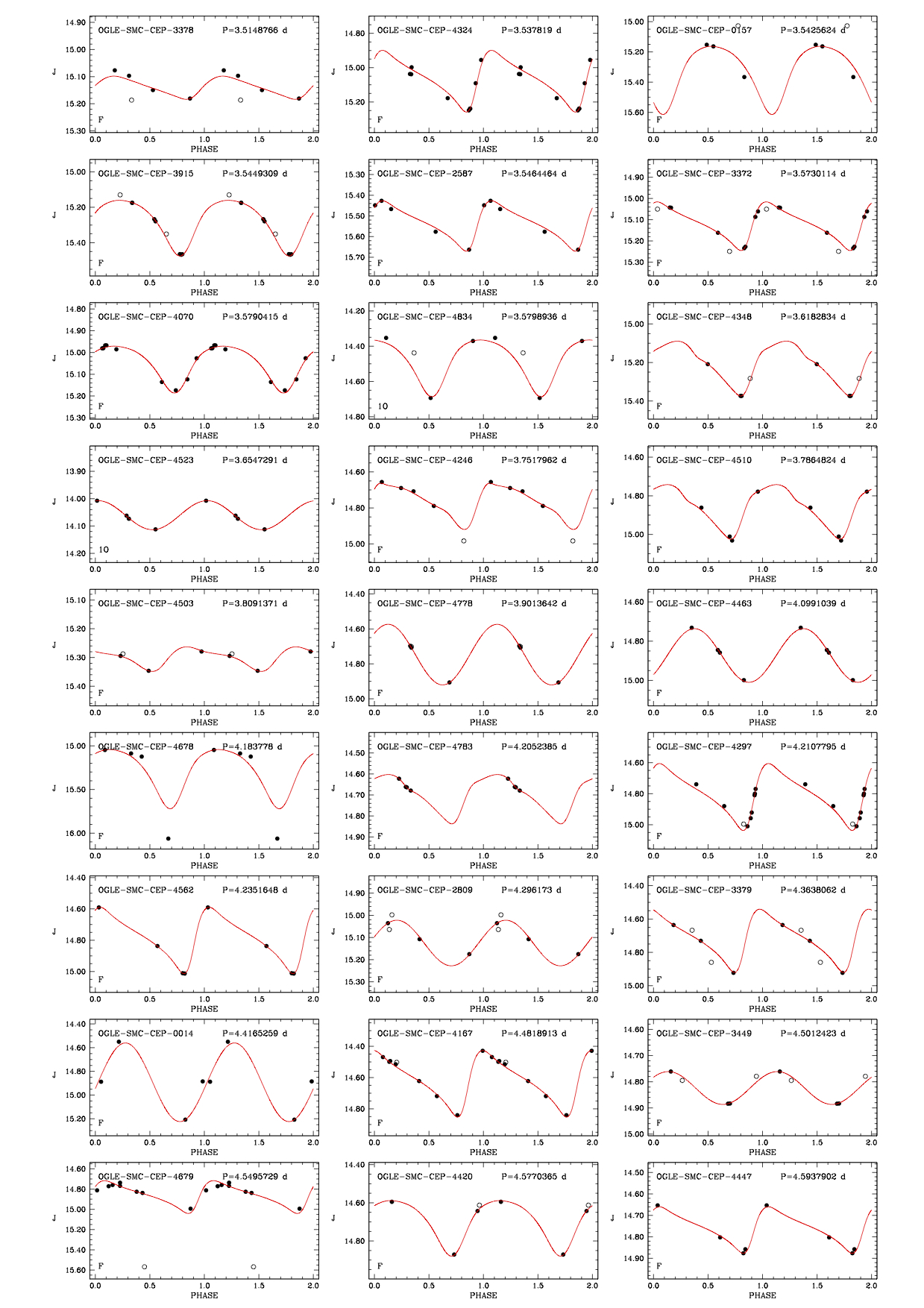}
\caption{Same as in Fig.~\ref{fig:CLY} but for the $J$-band.} 
\label{fig:CLJ}
\end{figure*}

\begin{figure*}
\includegraphics[width=16cm]{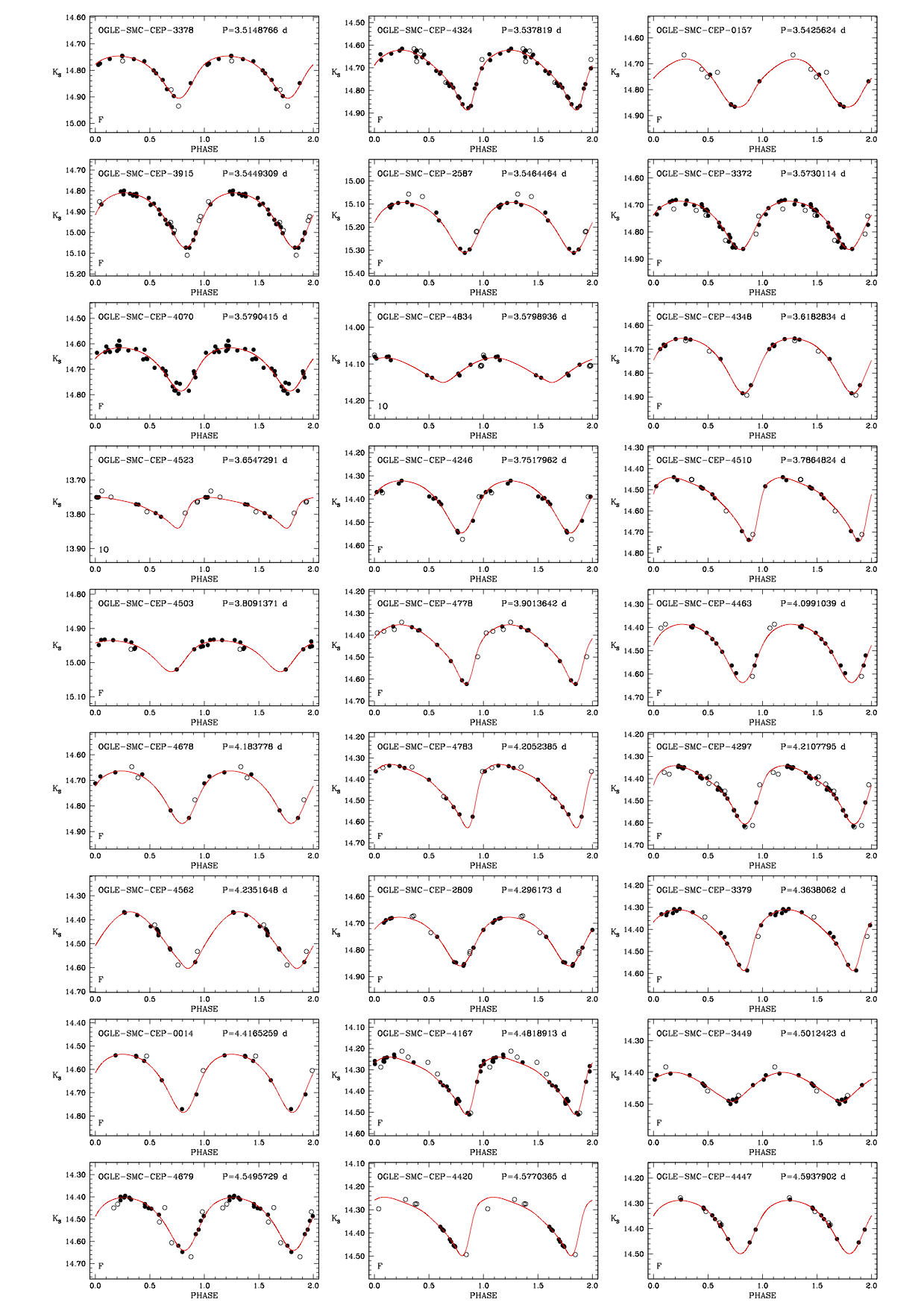}
\caption{Same as in Fig.~\ref{fig:CLY} but for the $K_\mathrm{s}$} 
\label{fig:CLK}
\end{figure*}

The VMC light curves in $Y,~J$ and $K_\mathrm{s}$  filters of  the 717 CCs analysed in this paper and not 
 present in Paper I are shown in Fig.~\ref{fig:CLY},  Fig.~\ref{fig:CLJ} and Fig.~\ref{fig:CLK}, respectively. The figures also 
 include the 40 stars already present in Paper I but for which the 
 number of epochs has been almost doubled. 

\section{Distances of ab-type RR Lyrae stars in SMC and LMC}
\label{appendixB}

To estimate the distances of the RR Lyrae stars in the LMC and SMC we 
followed substantially the same approach of \citet{Jacy2017}. In more 
detail, we used: i) the $V,I$ photometry provided by the OGLE\,IV survey 
\citep{Soszynski2016}; ii)  the theoretical $PW(V,I)$- [Fe/H]  
relations provided by \citet{Marconi2015} (their Table 
7). The usefulness of Wesenheit magnitudes has been already discussed 
in this paper and we will not repeat the same considerations 
here. \citet{Marconi2015} calculated $PW(V,I)$- [Fe/H]   relations for a variety of 
magnitude-colour combinations including $V,I$ bands: $W(V,I)=I-1.38(V-I)$. They list 
separately the relations for c- and ab-type RR Lyrae 
pulsators. However, both relations show a non-negligible dependence on  [Fe/H] : 

\begin{eqnarray}
W(V,I)=-0.94-2.43\log P+0.15{\rm [Fe/H]}~~~({\rm 0.03~mag})  \label{eqMarcella} \\
W(V,I)=-1.49-2.81\log P+0.13{\rm [Fe/H]}~~~({\rm 0.02~mag}) 
\end{eqnarray}

\noindent 
In the absence of spectroscopic  [Fe/H] measurements we can use the Fourier 
parameter $\phi_{31}$ (phase of the third harmonic minus three times 
the phase of the first harmonic) obtained from the Fourier 
decomposition of the light curve to estimate the metallicity \citep{Jurcsik1996}. To this 
end the ab-type RR Lyrae are more useful because they show a relation 
connecting $P$-$\phi_{31}$- [Fe/H]  which is better defined with 
respect to c-type pulsators \citep[see e.g.][]{Jurcsik1996,Nemec2013}. Therefore, we 
only considered ab-type RR Lyrae. They are abundant both 
in the LMC (25,926 objects) and SMC (4616 objects), hence statistics is 
not an issue. We choose to estimate the  [Fe/H]  value on the basis of 
the modern calibration of the $P$-$\phi_{31}$- [Fe/H]  relation of 
\citet{Nemec2013}. Their work is based on high-resolution 
spectroscopy and high-precision photometric light curves obtained with 
the Kepler satellite \citep{Nemec2011}. To transform $\phi_{31}$
in the $I$ band provided by OGLE\,IV into the $V$ band, required to use 
the \citet{Nemec2013} $P$-$\phi_{31}$- [Fe/H]   relation \cite[their $\phi_{31}$ is 
in the $K_p$ Kepler photometric band, which can however be easily 
transformed to the $V$-band using the conversion of][]{Nemec2011}, we 
adopted the following equation (Ripepi et al. in preparation): 

\begin{eqnarray}
\phi_{31}(V)=-(0.51\pm0.06)+(0.845\pm0.020)\phi_{31}(I)+ \\
 -(1.25\pm0.12)\log P~~({\rm 0.12~ rad}). \nonumber 
\end{eqnarray}

\noindent 
We used these values to estimate the ab-type RR Lyrae metallicity using 
the \citet{Nemec2013} results. In turn, these [Fe/H] in conjunction with 
pulsator periods from OGLE\,IV have been inserted in 
Eq.~\ref{eqMarcella} to obtain absolute $W(V,I)$ magnitudes. A simple 
comparison with the observed $W(V,I)$ calculated from OGLE\,IV 
photometry, directly provided the dereddened distance modulus, and in 
turn the distance, for each ab-type RR Lyrae in the LMC and SMC. The 
derived distances have been adjusted to provide an average distance of 
63.0 kpc for the SMC.  We remind that the detailed structure of the SMC based 
on VMC observations of RR Lyrae stars is presented in \citet{Muraveva2017}.


\bsp	
\label{lastpage}
\end{document}